\documentclass[a4paper,twocolumn,11pt, published=2017-05-09]{quantumarticle}
    \pdfoutput =1
    
    \usepackage[numbers,sort&compress]{natbib}
    \usepackage{braket}
    \usepackage{soul}
    \usepackage[utf8]{inputenc}
    \usepackage[english]{babel}
    \usepackage{dcolumn}
    \usepackage{bm}
    \usepackage[colorlinks=true,citecolor=blue,linkcolor=blue,urlcolor=blue,pdfencoding=auto]{hyperref}
    \usepackage[normalem]{ulem}
    \usepackage{acronym}
    \usepackage{color} 
    \usepackage{amssymb}
    \usepackage{mathrsfs}
    \usepackage{xcolor}
    \usepackage{subeqnarray}
    \usepackage{diagbox}
    \usepackage{makecell}
    \usepackage{mathtools}
    \usepackage{empheq}
    \usepackage{comment}
    \usepackage{bbm}

    \DeclareMathOperator{\sech}{sech}

    \newcommand{\bsb}[1]{\boldsymbol{#1}}

    \newcommand{\blambda}{\bsb{\lambda}}

    \newcommand{\Lc}{\mathcal{L}}

    \definecolor{gold}{RGB}{255, 100, 0}
    \definecolor{aprcolor}{rgb}{0.7,0.1,0.2}

    \newacro{GOAT}{Gradient Optimization of Analytic Controls}
    \newacro{QED}{Quantum Electrodynamics}
    \newacro{TMS}{Two-Mode Squeezing}
    \newacro{CD}{counterdiabatic}
    \newacro{STA}{Shortcuts to Adiabaticity}
    \newacro{LM}{Leakage Minimization}
    \newacro{QOC}{Quantum Optimal Control}
    \newacro{EL}{Euler-Lagrange}
    \newacro{QGT}{quantum geometric tensor}
    
\begin{document}


    \title{Geodesic steering for rapid dissipative entanglement distribution}

    \author{K. Venkataraman}
    \affiliation{Technical University of Munich, TUM School of Natural Sciences, Physics Department, 85748 Garching, Germany}
    \affiliation{Walther-Meißner-Institut, Bayerische Akademie der Wissenschaften, 85748 Garching, Germany}
    \affiliation{Munich Center for Quantum Science and Technology (MCQST), 80799 Munich, Germany}
    
    \author{P. Martinez-Azcona}
    \affiliation{Technical University of Munich, TUM School of Natural Sciences, Physics Department, 85748 Garching, Germany}
    \affiliation{Walther-Meißner-Institut, Bayerische Akademie der Wissenschaften, 85748 Garching, Germany}
    \affiliation{Munich Center for Quantum Science and Technology (MCQST), 80799 Munich, Germany}
    
    \author{M. Ljubotina}
    \affiliation{Technical University of Munich, TUM School of Natural Sciences, Physics Department, 85748 Garching, Germany}
    \affiliation{Munich Center for Quantum Science and Technology (MCQST), 80799 Munich, Germany}

    \author{A. Parra-Rodriguez}
    \affiliation{Technical University of Munich, TUM School of Natural Sciences, Physics Department, 85748 Garching, Germany}
    \affiliation{Walther-Meißner-Institut, Bayerische Akademie der Wissenschaften, 85748 Garching, Germany}
    \affiliation{Munich Center for Quantum Science and Technology (MCQST), 80799 Munich, Germany}
    
    \author{P. Rabl}
    
    \affiliation{Technical University of Munich, TUM School of Natural Sciences, Physics Department, 85748 Garching, Germany}
    \affiliation{Walther-Meißner-Institut, Bayerische Akademie der Wissenschaften, 85748 Garching, Germany}
    \affiliation{Munich Center for Quantum Science and Technology (MCQST), 80799 Munich, Germany}

    \date{11/11/2026}
    
    \begin{abstract}
    We describe an optimal-control strategy designed to accelerate the dissipative preparation of complex, highly entangled quantum many-body states. Analogous to optimal steering of isolated quantum systems, our scheme identifies control pulses that guide the evolution of open quantum systems along trajectories of instantaneous steady states while minimizing leakage resulting from non-adiabatic transitions. In many cases of interest, these control parameters can be derived analytically and without the need for computationally expensive numerical simulations of the full many-body master equation. We apply this technique for the preparation of bi- and multipartite entangled states in qubit systems coupled to a common photonic waveguide or a distributed squeezed reservoir. In both scenarios, our optimal-control strategy significantly outperforms the bare relaxation dynamics after a quench, reducing preparation times of near-maximally entangled states by orders of magnitude. This drastic speedup makes dissipative entanglement distribution protocols compatible with high-fidelity quantum information processing applications and illustrates more broadly the potential of leakage minimization for accelerating reservoir-engineering schemes. 
    \end{abstract}
    
    \maketitle
    
    
    \section{Introduction}

    While dissipation typically degrades quantum mechanical features such as superpositions and entanglement, \emph{quantum reservoir engineering} turns the coupling to an environment into a resource by stabilizing non-trivial quantum states through tailored dissipative processes. Initially proposed as a method for preparing squeezed motional states of a single trapped ion~\cite{Poyatos_96}, this concept has since been extended to the dissipative stabilization of far more complex many-body quantum states~\cite{Diehl_08, Kraus_08, Barreiro_11, Diehl_11, Muller_12, Schindler_13, Morigi_15, Cian_19} and explored as an alternative approach for a variety of tasks in quantum information processing~\cite{Verstraete_09,Kapit_17,Harrington_22}. Of particular interest in this context are purely dissipative schemes for the preparation and stabilization of highly entangled two- and multi-qubit states~\cite{Benatti_03}. Such states can be generated, for example, by coupling the qubits to a common lossy cavity mode~\cite{Wang_10, Kastoryano_11,  Rao_14, Aron_14, Reiter_16}, but also in extended settings where multiple qubits interact dissipatively via optical waveguides~\cite{Muschik_11, Stannigel_12, Li2012, Ramos_14, Pichler_15, Agusti_25} or distributed squeezed reservoirs~\cite{Kraus_04, Paternostro_04, Zippilli_13, Felicetti_14, You_18, Didier_18, Ma_19, Pocklington_22, Govia_22, Agusti_22, Agusti_23}. A key advantage of these dissipative protocols is that the desired entangled state emerges autonomously as the system's unique steady state, as demonstrated experimentally in a variety of platforms, including trapped ions~\cite{Lin_13, Cole_22}, atomic ensembles~\cite{Krauter_11}, and superconducting qubits~\cite{Shankar_13,Brown_22,Shah_24,Irfan_25,Yang_26,Juanes_26,Almanakly_26}.

    Despite these proof-of-concept demonstrations, a key limitation of dissipative entanglement distribution schemes is often that the relaxation time grows rapidly with the complexity of the target state, particularly with its degree of entanglement. Indeed, Pocklington and Clerk~\cite{Pocklington_24} have shown that under rather generic conditions, there exists a fundamental tradeoff between the time it takes to prepare the state and its final steady-state entanglement. In practical settings, where competing environmental loss channels like qubit dephasing impose strict time constraints, this bottleneck has hindered the use of reservoir engineering for generating highly entangled states so far. While including direct qubit-qubit interactions~\cite{Lim_24}, extending the Hilbert space~\cite{Didier_18,Yao_26} or employing measurements and feedback control~\cite{Riste_13,Roch_14,Andersen_19,Pocklington_25} can in principle overcome this tradeoff, doing so compromises the inherent simplicity or the distributed nature of the original scheme.

    In this article, we introduce and analyze an alternative approach that largely retains the simplicity of the original autonomous state-preparation schemes but produces near-maximally entangled states significantly more quickly by actively steering the system towards them. This approach is motivated by related ideas for optimal steering of isolated systems~\cite{Rezakhani_09,Tomka_16,Ljubotina_22} and evolves an open quantum system along a path of instantaneous steady states of the Liouville operator that describes the system's dissipative dynamics. This path is parametrized by one or a few external control parameters and adiabatically~\cite{Avron_12,Venuti_16} connects a trivial initial state with the complex entangled steady state of interest. By minimizing leakage~\cite{Ljubotina_22} out of this sub-manifold, we obtain a variational principle to optimally steer the system along this path, even at finite sweep times. 
    For families of pure steady states, the optimal control parameters are geodesics of the quantum metric tensor~\cite{Provost_80}, similar to what has been observed in the steering of fully isolated quantum systems~\cite{Tomka_16, Rezakhani_09, Dengis_26}. However, leakage minimization (LM)~\cite{Ljubotina_22} also naturally accounts for additional external control fields to implement counterdiabatic (CD) driving techniques~\cite{Demirplak_2003, Berry_09, Campo_13, Sels_17, Cepaite_23, Kolodrubetz_17, Duncan_24, Grabarits_26} and therefore provides a convenient unified framework for optimizing dissipative state preparation schemes.
    
    While LM does not guarantee globally optimal control, it is an \emph{informed} approach that leverages the known analytical structure of the system's steady state. Here we show how this property can be exploited to boost the performance of entanglement distribution schemes in two common settings where multiple qubits are coupled to (i) a single waveguide and (ii) a two-mode-squeezed (TMS) photonic reservoir. In both scenarios, there exist analytically known families of bi- and multipartite entangled dark states, which are pure steady states of the Liouvillian for a fixed set of driving parameters~\cite{Stannigel_12, Pichler_15, Agusti_23}. However, the time it takes for the system to relax into these states increases substantially with the degree of entanglement of the final steady state, in accordance with the time-entanglement tradeoff identified in Ref.~\cite{Pocklington_24}. By applying LM to a single control parameter, we obtain optimized control pulses that can reduce the preparation time by more than two orders of magnitude in cases where the target states are close to maximally entangled. This massive speedup makes dissipative entanglement preparation schemes compatible with quantum information processing applications, where stringent fidelity requirements for entangling operations are still out of reach for existing protocols.
    
    Importantly, the current analysis shows that in many cases of interest, LM control pulses can be obtained either analytically or by numerically solving simple differential equations for a small set of parameters. This crucial property makes the method applicable for large multi-qubit systems, where standard quantum optimal control (QOC) techniques---which rely on integrating the full master equation---become numerically intractable. Even in regimes where full-fledged optimal control remains feasible, we find that LM yields pulses that are already near-optimal. As an illustrative example, we also discuss the combination of LM with additional local CD control fields. While---as expected---local control provides only minor improvement for generating highly entangled states, it can still speed up the distribution of states with low or moderate entanglement. These findings confirm previous observations~\cite{Ljubotina_22} for closed quantum systems, where LM outperforms other common variational methods~\cite{Sels_17,Claeys_19,Grabarits_26, Cepaite_23, Duncan_25, Hegade_21_PRA, Hegade_21, Hegade_22} for CD control. It can thus provide similar performance gains for a variety of dissipative state-preparation schemes \emph{without} requiring detailed knowledge of the full Liouvillian spectrum~\cite{Santos_21, Vacanti_2014, Alipour_2020, Santos_21_TQD} or an exact solution for the time-evolving density matrix \cite{Dann_19, Dupays_20}.

    The remainder of the paper is structured as follows. In Sec.~\ref{sec: LM_for_Diss_sys} we describe the general formalism of LM and its connections to the previous literature. In Sec.~\ref{sec: DD_ent_gen} and Sec.~\ref{sec: TMS_ent_gen}, we then describe the application of this formalism for dissipative entanglement distribution in systems of qubits coupled to one-dimensional waveguides and TMS reservoirs, respectively. Finally, in Sec.~\ref{sec: QOC_CD} we benchmark LM against full QOC and investigate the application of additional CD fields. The main findings of this work are summarized in Sec.~\ref{sec: Conclusions}.

    \section{Leakage minimization for dissipative quantum systems}
    \label{sec: LM_for_Diss_sys}
    
    \begin{figure}[t]
        \centering
        \includegraphics[width=\linewidth]{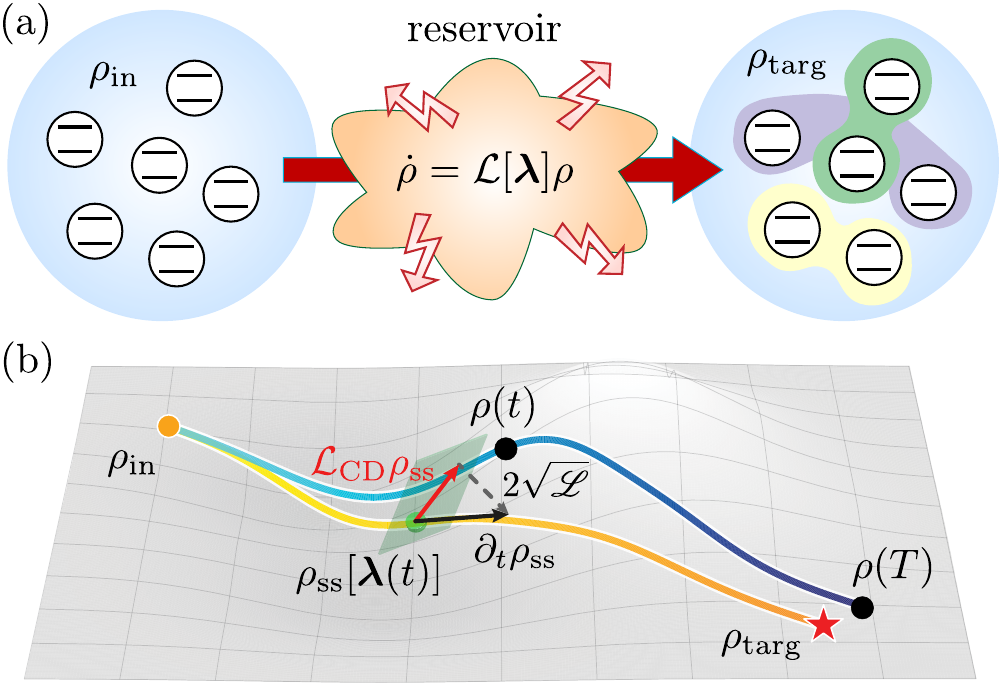}
        \caption{(a) Illustration of reservoir engineering. Starting from a simple initial state $\rho_{\rm in}$, a quantum many-body system can be dissipatively driven into a complex, highly entangled state $\rho_{\rm targ}$, which is the steady state of an appropriately engineered Liouvillian dynamics. (b) Schematic of the LM approach for steering dissipative quantum systems along a trajectory (orange path) of instantaneous steady states, $\rho_{\rm ss}[\boldsymbol{\lambda}(t)]$, which are parametrized by a set of control parameters $\boldsymbol{\lambda}(t)$.  At each point in time, the loss function $\mathscr{L}$ accounts for the squared difference between the infinitesimal change along this parametrized curve, $\partial_t \rho_{\rm ss}$, and the evolution generated by the system's total Liouvillian, $\mathcal{L}_{\rm tot}\rho_{\rm ss}$. This evolution is trivial when $\mathcal{L}_{\rm tot}=\mathcal{L}$, but contributes a term $\mathcal{L}_{\rm CD}\rho_{\rm ss}$ (indicated by the red arrow), when additional CD control fields are applied to counteract the nonadiabatic corrections.  As long as the integrated losses are sufficiently small, the evolution of the actual quantum state $\rho(t)$ (blue path) will follow the trajectory of instantaneous steady states closely and reach the targeted state $\rho_{\rm targ}\equiv \rho_{\rm ss}[\boldsymbol{\lambda}(T)]$ at the final time $T$. See text for more details.}
        \label{fig:Schematic}
    \end{figure}
    
    We consider a generic dissipative quantum many-body system whose dynamics is governed by a Markovian master equation,
    \begin{align}
        \partial_t \rho(t) = \Lc \rho(t). 
        \label{eq: ME}
    \end{align}
    Here, the density operator $\rho(t)$ represents the state of the quantum system at time $t$ and $\Lc\equiv \Lc[\boldsymbol{\lambda}]$ is the Liouville operator, which depends on a set of control parameters $\boldsymbol{\lambda}=\{\lambda_1,\lambda_2, \dots\}$. For a time-independent Liouvillian $\Lc$ with $\boldsymbol{\lambda}\neq \boldsymbol{\lambda}(t)$, there exists at least one steady state satisfying
    \begin{align}\label{eq:SteadyStateGeneral}
            \Lc[\boldsymbol{\lambda}] \rho_\text{ss}[\boldsymbol{\lambda}] = 0.
    \end{align}
    If the steady state is unique, which we assume throughout this work, any initial state $\rho_{\rm in}=\rho(t=0)$ will approach $\rho_\text{ss}[\boldsymbol{\lambda}]$ for sufficiently long times. Interestingly, for an appropriately chosen $\Lc$, the resulting steady states can be pure and represent complex and highly entangled quantum many-body states \cite{Stannigel_12, Pichler_15, Diehl_08, Barreiro_11, Muller_12, Schindler_13}. Consequently, such states can be prepared fully autonomously through a dissipative coupling to an engineered reservoir, as illustrated in Fig.~\ref{fig:Schematic}(a).

    While Eq.~\eqref{eq:SteadyStateGeneral} ensures that the system eventually approaches $\rho_\text{ss}[\boldsymbol{\lambda}]$, in many cases of interest, one finds that the time to prepare the steady state (or to come sufficiently close to it) increases with the system size \cite{Agusti_23} or with the degree of entanglement~\cite{Pocklington_24}. This tradeoff limits the ability to dissipatively prepare complex many-body states in realistic experimental settings, where the engineered dissipation processes compete with other unavoidable decoherence mechanisms.

    \subsection{Time-dependent control for  dissipative state preparation}
    
    To overcome this bottleneck, we consider a more general scenario in which we allow for time-dependent control parameters, $\boldsymbol{\lambda}\equiv \boldsymbol{\lambda}(t)$, in order to accelerate the relaxation dynamics. Specifically, starting from a  trivial initial steady state $\rho_{\rm in}\equiv\rho_\text{ss}[\boldsymbol{\lambda}(0)]$ at time $t=0$, our goal is to identify a set of optimized control pulses $\boldsymbol{\lambda}(t)$, such that the time-evolved state $\rho(T)$ approaches the targeted steady state at a minimal final time $t=T$, i.e.,
    \begin{equation}
    \rho(T) \simeq \rho_{\rm targ} \equiv \rho_\text{ss}[\boldsymbol{\lambda}(T)].
    \end{equation}
    To quantify the outcome of such a protocol, we introduce the fidelity
    \begin{align}
       \mathcal{F}\left(\rho_1|\rho_2\right)=\Big( \mathrm{Tr}\Big\{\sqrt{\sqrt{\rho_1} \rho_2 \sqrt{\rho_1}}\Big\}\Big)^2,
      \label{eq:fidelity}
    \end{align}
    which reduces to the state overlap, $\mathcal{F}=\bra{\psi_1}\rho_2\ket{\psi_1}$, when $\rho_1=|\psi_1\rangle\langle \psi_1|$ is pure. The success of preparing the state $\rho_{\rm targ}$ can then be formulated in terms of the fidelity surpassing a certain threshold, i.e., 
    \begin{align}
       \mathcal{F}\left(\rho(T)|\rho_{\rm targ}\right)\geq  1- \zeta,
    \end{align}
    where the residual tolerable error $\zeta$ depends on the specific application. We are interested in minimizing the time $T$ at which this threshold is first reached. 
    
    This general type of problem can in principle be solved with the help of optimal control techniques \cite{Machnes_18,Horn2018,Koch_22,Chen_25,Ding_26}, which have been widely used in the literature for controlling the coherent or dissipative evolution of quantum systems. These techniques, however, become numerically very expensive and inefficient when either the number of control parameters or, especially, the system size increases. In particular, in the case of dissipative quantum many-body systems, where the master equation must be numerically integrated during the optimization process many times, such brute-force optimization techniques become quickly intractable \cite{Lu_24, Larocca_2022, Moore_11, Fentaw_25, Metz_2023}. 
    
    \subsection{Optimal steering of dissipative quantum systems} 
    
    Here we introduce an \emph{informed} optimization technique, which is inspired by related optimal steering schemes for isolated quantum systems~\cite{Tomka_16,Ljubotina_22} and becomes particularly useful for reservoir-engineering applications. Here, the instantaneous steady state satisfying Eq.~\eqref{eq:SteadyStateGeneral} is analytically known and there is a parameter set $\boldsymbol{\lambda}(0)$, for which the corresponding steady state is simple and experimentally easy to prepare. Therefore, 
    for a given continuous path $\boldsymbol{\lambda}(t)$ in parameter space, we obtain a corresponding path in the space of density operators,
    \begin{align}\label{eq:path_density_operator}
     \rho_\text{ss}[\boldsymbol{\lambda}(0)] \longrightarrow \rho_{\rm ss}[\boldsymbol{\lambda}(t)] \longrightarrow \rho_{\rm ss}[\boldsymbol{\lambda}(T)].
    \end{align}
    As illustrated in Fig.~\ref{fig:Schematic}(b), this path smoothly interpolates between the initial state $\rho_{\rm in}$ and the targeted state $\rho_{\rm targ}$. The actual state of the system, $\rho(t)$, will adiabatically~\cite{Avron_12} follow this path, as long as the parameters are varied sufficiently slowly. The objective of the proposed steering technique is to enforce an evolution of the true quantum state $\rho(t)$ that closely follows the path in Eq.~\eqref{eq:path_density_operator}, even when the parameters $\blambda$ are varied at finite speed. 
    
    To this end, we consider a generic family of density operators $\varrho_t$, parametrized by time $t$, together with the loss function
    \begin{align}\label{eq:loss_general}
            \mathscr L(t) = \frac{1}{4}\Big\lVert  \partial_t \varrho_t -\Lc[\boldsymbol{\lambda}(t)]\varrho_t \Big\rVert^2,
    \end{align}
    where $\lVert A\rVert  =  \text{Tr}\{\sqrt{{A^\dag A}}\}$ is the trace norm of the operator $A$. This function measures the difference between the state that is evolved (for a small time step) according to its parametrization and the actual evolution of the same state under the system's Liouvillian. By identifying $\varrho_t\equiv \rho_{\rm ss}[\boldsymbol{\lambda}(t)]$ with the path of instantaneous steady states  
    in Eq.~\eqref{eq:path_density_operator}, the second term inside the norm vanishes, and we obtain the simple result
    \begin{align}
        \mathscr L(t)&=  \frac{1}{4}\Big\lVert \partial_t  \rho_{\rm ss}[\boldsymbol{\lambda}(t)] \Big\rVert^2  \nonumber \\
        & = \lim_{dt\to
        0} \frac{1}{4}\frac{\left \lVert\rho_{\rm ss}[\boldsymbol{\lambda}(t+dt)]- \rho_{\rm ss}[\boldsymbol{\lambda}(t)]\right\rVert^2}{dt^2}. \nonumber
     \end{align} 
    The condition $\mathscr L(t)=0$ for $t\in[0,T]$ implies that the actual evolution of the system described by $\rho(t)$ coincides with the parametrized path $\rho_{\rm ss}[\boldsymbol{\lambda}(t)]$ and we reach $\rho_{\rm targ}$ with perfect fidelity. 
    
    In most situations, this condition will not be satisfied exactly, but as long as $\mathscr L(t)$ remains sufficiently small at all times, $\rho_{\rm targ}$ can be reached with high accuracy. Therefore, we are interested in control pulses $\boldsymbol{\lambda}(t)$ that minimize 
    the total leakage error,
    \begin{align}
       \mathscr{E}_L = \int_0^T \mathscr L\left(\boldsymbol{\lambda}(t), \dot{\boldsymbol \lambda}(t)\right) dt.
        \label{eq: Var_act}
    \end{align}
    This quantity is minimized by control parameters $\boldsymbol{\lambda}(t)$ that satisfy the Euler-Lagrange equations
    \begin{align}
        \begin{split}
        \frac{d}{dt}\left(\dfrac{\partial \mathscr L}{\partial\dot{\lambda}_i}\right) = \dfrac{\partial \mathscr L} {\partial \lambda_i},
        \label{eq: EL_Eq}
    \end{split}    
    \end{align}
    which are labeled by $\boldsymbol{\lambda}^{\rm LM}(t)$ in the following. We emphasize that this set of equations only involves a small number of control parameters and can be derived without the explicit knowledge of the full Liouvillian's spectrum. 
    
    \subsection{Pure states and geodesic control}\label{subsec:Pure_States_Geodesic}
    In many cases of interest, the instantaneous steady states are pure, i.e.,  $\rho_{\rm ss}[\boldsymbol{\lambda}]= |\psi_D^{\boldsymbol{\lambda}}\rangle\langle \psi_D^{\boldsymbol{\lambda}}|$. Such a pure 
     steady state $|\psi_D^{\boldsymbol{\lambda}}\rangle$ is typically referred to as a \emph{dark state},  which is an eigenstate of the Hamiltonian and annihilated by all jump operators in the Liouvillian~\cite{Kraus_08}. While the general framework described above does not rely on this assumption, it greatly simplifies the loss function (see Ref.~\cite{Nielsen_00} and Appendix~\ref{app:Loss_Pure_States})
    \begin{align}\label{eq:LossFunction_Pure}
       \mathscr L(t) =  \lim_{dt\to
        0}  \frac{1- \left|\braket{\psi_D^{\boldsymbol{\lambda}(t+dt)}|\psi_D^{\boldsymbol{\lambda}(t)}}\right|^2}{dt^2}.
    \end{align}
    Taking the limit $dt\rightarrow 0$, Eq.~\eqref{eq:LossFunction_Pure} can be written in a compact form as
    \begin{align}\label{eq:loss_multi_param}
        \mathscr L(\boldsymbol{\lambda}, \dot{\boldsymbol \lambda})=  \sum_{\mu, \nu} g_{\mu, \nu} \dot{\lambda}_\mu \dot{\lambda}_\nu =: g_{t,t}.
    \end{align}
    Here, 
    \begin{align}\label{eq:QGT}
     g_{\mu,\nu}=  {\rm Re}\left\{
    \partial_{\lambda_\mu}\bra{\psi_D^{\boldsymbol{\lambda}}} \left(\mathbbm{1}- \ket{\psi_D^{\boldsymbol{\lambda}}}\bra{\psi_D^{\boldsymbol{\lambda}}}\right)\partial_{\lambda_\nu}\ket{\psi_D^{\boldsymbol{\lambda}}}
    \right\},
    \end{align}
    is the so-called quantum metric tensor, as originally introduced by Provost and Vallee~\cite{Provost_80}. Correspondingly, $g_{t,t}$ can be interpreted as its time-component, or as the squared velocity along the state trajectory. 
    
    From Eq.~\eqref{eq:loss_multi_param} and Eq.~\eqref{eq:QGT} we now explicitly see that the loss function is only sensitive to the component of $\partial_\lambda\ket{\psi_D^{\boldsymbol{\lambda}}}$ orthogonal to $\ket{\psi_D^{\boldsymbol{\lambda}}}$. Therefore, it measures leakage of population out of the predefined path [see Fig.~\ref{fig:Schematic}(b)], but is insensitive to variations of the global phase of $\ket{\psi^{\boldsymbol{\lambda}}_D}$.  Since $g_{\mu,\nu}$ represents a Riemannian metric tensor on the manifold of control parameters, the optimized parameters $\boldsymbol{\lambda}^{\rm LM}(t)$ that minimize $\mathscr{E}_L$ also correspond to `geodesics' that minimize the total length \cite{Milnor_63, Docarmo_92}
    \begin{equation}
    L_{\rm path} 
    = \int_0^T dt\sqrt{g_{t,t}}
    \end{equation}
    of the path between states $\ket{\psi_D(\blambda(0))}$ and $\ket{\psi_D(\blambda(T))}$. Note, however, that the opposite is not true and different reparametrizations of the same minimal path do not minimize the leakage error. As shown in Appendix~\ref{app:L_Invariance}, from this geometric interpretation it also follows that the loss function is a constant of motion, i.e., 
    \begin{equation}
    \mathscr L\left(\boldsymbol{\lambda}^{\rm LM}(t), \dot{\boldsymbol{\lambda}}^{\rm LM}(t)\right)= \rm{const}.
    \end{equation}
    For the examples discussed below, this property is particularly useful to derive exact analytic expressions for the control parameters.

    \subsection{Counterdiabatic driving}
    
    So far we have considered a set of intrinsic control parameters, $\boldsymbol{\lambda}$, for which the corresponding path of steady states, $\rho_{\rm ss}[\boldsymbol{\lambda}]$, is known. However, leakage out of this path could potentially be further suppressed by applying additional external control fields to counteract any of the remaining nonadiabatic transitions. This approach is known as CD control and can be readily included in the LM framework by considering the total Liouvillian
    \begin{align}
         \Lc_{\rm tot}[\boldsymbol{\lambda},\boldsymbol{\beta}]  = \Lc[\boldsymbol{\lambda}]+ \Lc_{\rm CD}[\boldsymbol{\beta}].
         \label{eq: L_c_CD}
    \end{align}
    Here, $\Lc[\boldsymbol{\lambda}]\rho_{\rm ss}[\boldsymbol{\lambda}]=0$ as before and $\Lc_{\rm CD}[\boldsymbol{\beta}]$ accounts for additional CD control fields with time-dependent parameters $\boldsymbol{\beta}(t) = \{\beta_1(t), \beta_2(t),\dots\}$. 
    
    By replacing the original Liouvillian $\mathcal{L}$ that determines the system evolution in Eq.~\eqref{eq:loss_general} with the total Liouvillian $\mathcal{L}_{\rm tot}$, we obtain the generalized loss function 
    \begin{align}
        \mathscr L(t)&=  \frac{1}{4}\Big\lVert \partial_t  \rho_{\rm ss}[\boldsymbol{\lambda}(t)]- \Lc_{\rm CD}[\boldsymbol{\beta}(t)]\rho_{\rm ss}[\boldsymbol{\lambda}(t)] \Big\rVert^2 \nonumber\\
        &\equiv \mathscr L(\boldsymbol{\lambda},\dot{\boldsymbol{\lambda}},\boldsymbol{\beta}).
        \label{eq: Loss_CD}
     \end{align} 
    Since it does not depend on the velocities $\dot{\boldsymbol{\beta}}$, the Euler-Lagrange equations 
    \begin{align}
        \frac{\partial \mathscr{L}(\boldsymbol{\lambda}, \dot{\boldsymbol{\lambda}}, \boldsymbol{\beta})}{\partial\beta_i} = 0,
        \label{eq: EL_CD}
    \end{align}
    can be used to eliminate the variables $\boldsymbol{\beta}$ and derive an effective loss function $\mathscr{L}_{\rm eff}(\boldsymbol{\lambda}, \dot{\boldsymbol{\lambda}})$ for the original control parameters only. 
    
    For systems with a pure steady state and Hamiltonian driving, i.e.,  $\Lc_{\rm CD}[\boldsymbol{\beta}]\rho = - i[H_{\rm CD}[\boldsymbol{\beta}], \rho]$, the loss function reduces to (see Appendix~\ref{app:Loss_Pure_States})
    \begin{align}
        \mathscr{L}_{\rm eff} = \Big\lvert \Big(\mathbbm{1} -\ket{\psi_D^{\boldsymbol{\lambda}}}\bra{\psi_D^{\boldsymbol{\lambda}}}\Big)\Big(\partial_t +iH_{\rm CD}[\boldsymbol{\beta}]\Big)\ket{\psi_D^{\boldsymbol{\lambda}}} \Big\rvert^2.
        \label{eq:LossFunction_CD_Pure}
    \end{align}
    Here, $|\bullet|$ is the usual vector norm, $\boldsymbol{\lambda}\equiv \boldsymbol{\lambda}(t)$ and $\boldsymbol{\beta}\equiv \boldsymbol{\beta}(\blambda, \dot{\blambda})$ denotes the minimizing CD amplitudes. This expression is equivalent to the loss function proposed in Ref.~\cite{Ljubotina_22} for CD control of isolated systems. It can be readily cast into the general form in  Eq.~\eqref{eq:loss_multi_param} by introducing the modified quantum metric tensor
    \begin{align}\label{eq:QGT_CD}
     \bar g_{\mu,\nu}= {\rm Re} \left\{\partial_{\lambda_\mu}\bra{ \bar \psi_D^{\boldsymbol{\lambda}}}  \left(\mathbbm{1}- \ket{ \bar \psi_D^{\boldsymbol{\lambda}}}\bra{ \bar \psi_D^{\boldsymbol{\lambda}}}\right)\partial_{\lambda_\nu}\ket{ \bar \psi_D^{\boldsymbol{\lambda}}}\right\},
    \end{align}
    defined for the family of rotated states 
    \begin{equation}
    \ket{\bar\psi^{\boldsymbol{\lambda}}_D}
    = \mathcal{T}_-\exp\left(
    i\int_0^t H_{\rm CD}(t')\,dt'
    \right)
    \ket{\psi_D^{\boldsymbol{\lambda}(t)}}.
    \end{equation}
    More details about this derivation are provided in Appendix~\ref{app:Loss_Pure_States}.

    \subsection{Summary and relation to previous work}
    In conclusion, the LM approach introduced in this section provides a versatile framework for steering open quantum systems along a trajectory of instantaneous steady states. For the relevant case where steady states are pure, it recovers previous steering strategies for isolated quantum systems, which have been derived from the time-component of the quantum metric tensor along a path~\cite{Tomka_16} or from a LM ansatz for CD control~\cite{Ljubotina_22}. Interestingly, the discussion above shows that with the help of the effective metric in Eq.~\eqref{eq:QGT_CD}, these two seemingly different approaches can be unified into a single control framework with a common geometric interpretation and extended to open quantum systems. 
    
    Note that a geometric formalism, as developed in Sec.~\ref{subsec:Pure_States_Geodesic}, is not unique to the current setting. Similar formulations appear very naturally in a variety of quantum control problems~\cite{Tomka_16, Dengis_26, Rezakhani_09, Santos_21,Bukov_19, Campo_12, Goll_26}, where details of the underlying system and optimization strategy are contained in the respective metric tensor.  
    In the context of reservoir engineering, the crucial feature of LM is that despite optimizing the evolution of a dissipative quantum system, the resulting control problem only involves the quantum geometric tensor associated with the family of steady states, which is often known analytically. The structure of the full Liouvillian only affects the optimization implicitly through the condition in Eq.~\eqref{eq:SteadyStateGeneral}. This represents a crucial advantage over brute-force numerical optimization as well as shortcuts to adiabaticity which---for dissipative systems---require a detailed knowledge of the full Liouvillian spectrum \cite{Santos_21, Vacanti_2014, Alipour_2020, Santos_21_TQD}. 
    
    

    \section{Rapid entanglement distribution in a 1D quantum network}
    \label{sec: DD_ent_gen}
    \begin{figure}
        \centering
       \includegraphics[width=\columnwidth]{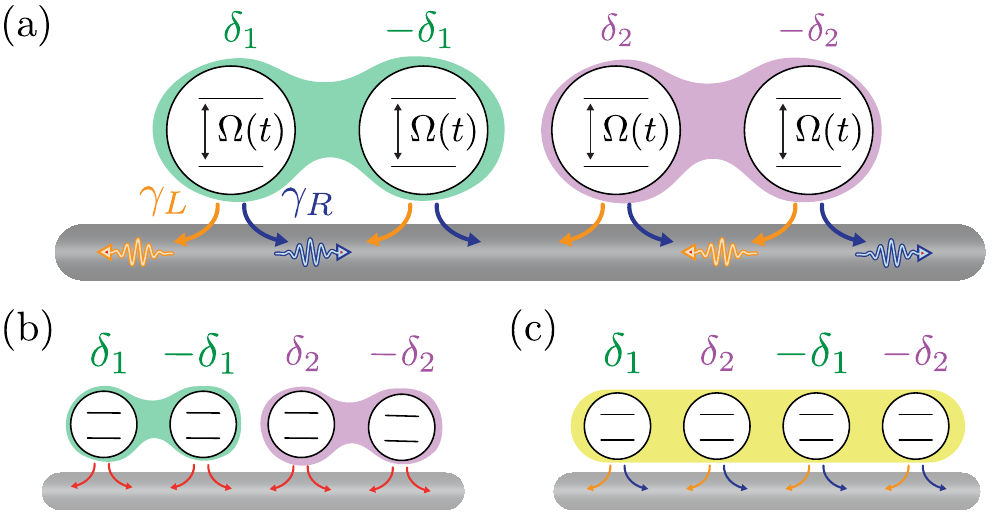} 
        \caption{(a) Schematic of the waveguide QED setup considered in Sec.~\ref{sec: DD_ent_gen}. Multiple qubits are driven by a classical field of strength $\Omega(t)$ and decay into the right- and left-propagating modes of a common 1D waveguide with rates $\gamma_R$ and $\gamma_L$, respectively.  For pairwise anti-matched detunings, the steady state of this system is a pure state of the general form given in Eq.~\eqref{eq: Casc_Dim_Dark}, which can be (b) pairwise entangled for alternating detunings or bidirectional waveguides or (c) multipartite entangled when $\gamma_R\neq \gamma_L$ and the detunings are permuted. See text and Refs.~\cite{Stannigel_12,Pichler_15} for more details.}
        \label{fig: Pichler_Schematic}
    \end{figure}
    
    As a first example, we consider the dissipative preparation of entangled steady states in a chain of $N$ driven two-level systems (qubits) that are coupled to a common photonic waveguide [see Fig.~\ref{fig: Pichler_Schematic}(a)]. The qubits with ground state $|g\rangle$ and excited state $|e\rangle$ are strongly driven locally by a coherent field and exchange photons through the waveguide. For this setup, it has been shown~\cite{Stannigel_12} that a continuous coherent driving of all qubits, combined with the collective, waveguide-mediated dissipation, can support the formation of pure, highly entangled steady states for an arbitrary number of qubits. For unidirectional or chiral waveguides, these steady states can exhibit genuine multipartite entanglement with an entanglement structure that is determined by the pattern of qubit detunings. 
    
    Following the more detailed analysis of this system presented in Refs.~\cite{Stannigel_12, Pichler_15}, we model this setup by a generalized cascaded master equation,
    \begin{align}
    \begin{split}
        \dot{\rho}(t) = &-i\Big[ \sum_{j=1}^N \left(\Omega_j(t) \sigma^x_j + \delta_j\sigma^+_j \sigma^-_j\right) + H_\text{casc}, \rho(t) \Big] \\
        &+  (\gamma_L+\gamma_R)\mathcal{D}[\Sigma]\rho(t),
        \label{eq: Pichler_ME}
    \end{split}
    \end{align}   
    where $\rho(t)$ denotes the $N$-qubit density operator and the $\sigma_j^k$ are the usual Pauli operators. In Eq.~\eqref{eq: Pichler_ME}, we have introduced the shorthand notation $\mathcal{D}[c]\rho= c\rho c^\dag - \{c^\dag c,\rho\}/2$ and $\Sigma = \sum_{j=1}^N \sigma^-_j$ is the collective lowering operator.
    The first line of Eq.~\eqref{eq: Pichler_ME} accounts for the coherent evolution and describes a set of individually driven qubits with driving strengths $\Omega_j(t)$ 
    and detunings $\delta_j$. The additional interaction, 
    \begin{align} H_\text{casc} =  - i \frac{\gamma_A}{2}\sum_{k>j}\left(\sigma^+_k \sigma^-_j - \text{H.c.}\right)
    \end{align}
    arises from an asymmetry $\gamma_A=\gamma_R-\gamma_L$ between the emission into right- and left-propagating waveguide modes with rates $\gamma_R$ and $\gamma_L$, respectively. This contribution vanishes in a bidirectional waveguide with $\gamma_R = \gamma_L$. For further details about the derivation of Eq.~\eqref{eq: Pichler_ME} and recent experimental realizations of this setup, the reader is referred to Refs.~\cite{Stannigel_12, Pichler_15, Shah_24,Irfan_25,Yang_26,Almanakly_26}.

    \subsection{Two-qubit entanglement}
    To illustrate the LM method, we first consider the simplest case of $N=2$ qubits coupled to a  waveguide. 
    We assume opposite detunings, $\delta_1=-\delta_2=\delta$, and equal driving strengths, $\Omega_i=\Omega$. Under these assumptions, the master equation in Eq.~\eqref{eq: Pichler_ME} has a pure steady state, $\rho_{\rm ss}=|\psi_D\rangle\langle \psi_D|$, where
    \begin{equation}
    \begin{split}
        |\psi_D\rangle= |D(\alpha)\rangle \equiv \frac{|gg\rangle - \alpha|S\rangle}{\sqrt{1+ |\alpha|^2}}.
        \label{eq: Dark_2qbit}
    \end{split}
    \end{equation}
    Here, $|S\rangle=(\ket{eg} - \ket{ge})/\sqrt{2}$ denotes the singlet state and $\alpha = \sqrt{2}\Omega/(\delta-i\gamma_A/2)$. Thus, the system relaxes into an entangled and pure two-qubit state, which approaches the maximally entangled singlet state in the limit $|\alpha| \rightarrow \infty$.

    \subsubsection{Relaxation after a quench}
\begin{figure}[t]
        \centering
        \includegraphics[width=\columnwidth]{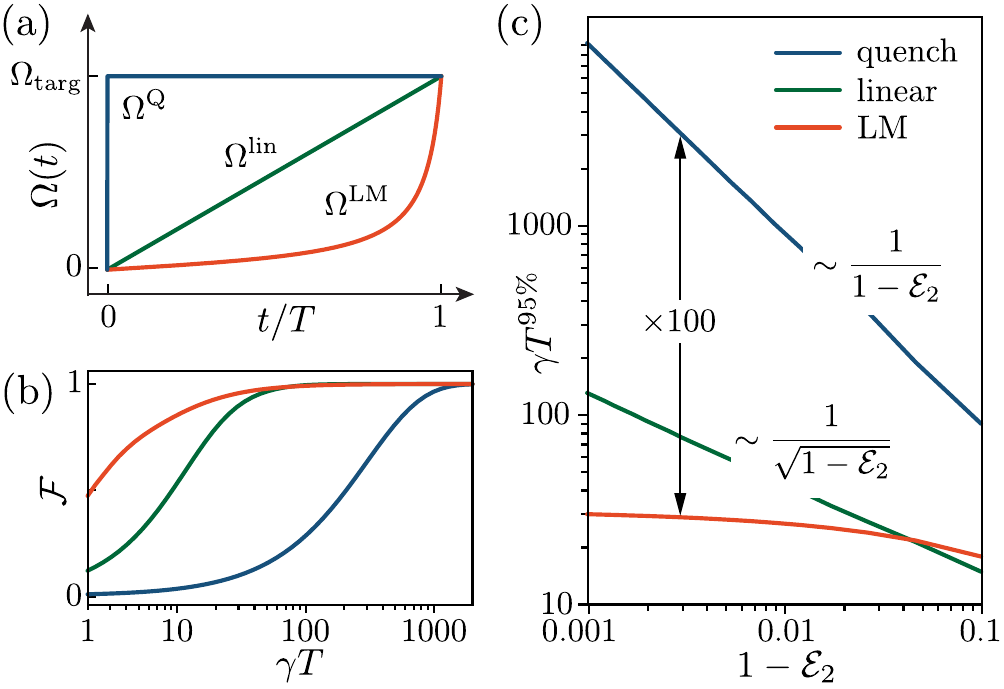}
        \caption{(a) Sketch of the three different sweep functions for the control parameter $\Omega$ assumed in the analysis of Sec.~\ref{sec: DD_ent_gen}. For these three control schemes, the resulting state preparation fidelity $\mathcal{F}$ is plotted in (b) as a function of the protocol time $T$. For this plot, a bidirectional waveguide with $\gamma_A=0, \delta = 2\gamma$, and $\Omega_{\mathrm{targ}} = 10\delta$ has been assumed, which corresponds to a targeted entanglement of $\mathcal{E}_2=0.99$. 
        (c) The minimal time $T^{95\%}$ that is required to reach a fidelity of $\mathcal{F}\geq 0.95$ is plotted against the entanglement $\mathcal{E}_2$ of the targeted steady state over a range $\Omega_{\rm targ}\in [3.05\delta, 32\delta]$. The other parameters are the same as in (b).}
        \label{fig: 2Qbit_Pichler}
    \end{figure}

    A simple way to prepare the state in Eq.~\eqref{eq: Dark_2qbit} with a given parameter $\alpha_{\rm targ}=\sqrt{2}\Omega_{\rm targ}/(\delta-i\gamma_A/2)$ is to initialize the qubits in state $|gg\rangle$ at time $t=0$ and then switch the driving field to its maximal value. This quench corresponds to the control pulse [see Fig.~\ref{fig: 2Qbit_Pichler}(a)] 
    \begin{equation}
    \Omega^{\text{Q}}(t) = 
    \begin{cases} 
       0 & \text{for }  t\leq 0 \\[10pt]
       \Omega_\text{targ} & \text{for }  t\in (0,T]  
    \end{cases}
    \end{equation}
    and represents the usual scenario considered in the analysis of dissipative state-preparation schemes. Naively, this approach also seems optimal, as relaxation toward the desired final steady state starts immediately.

    In Fig.~\ref{fig: 2Qbit_Pichler}(b) we plot the resulting state preparation fidelity $\mathcal{F}\left(\rho(T)|\rho_{\rm targ}\right)$, where $\rho_{\rm targ}=|D(\alpha_{\rm targ})\rangle\langle D(\alpha_{\rm targ})| $, as a function of the waiting time $T$ after the quench. 
    For large $|\alpha_{\rm targ}|$ and long enough protocol times, we find that the targeted state is approached exponentially, 
    \begin{equation}
    \label{eq: Fid_exp_Pichler}
    \mathcal{F} \simeq 1-e^{-\gamma_{\rm rel} T},
    \end{equation} 
    with a relaxation time, $\tau_{\rm rel}=\gamma^{-1}_{\rm rel}$. This time is considerably longer than the characteristic decay time, $\gamma^{-1}$, where $\gamma=(\gamma_L+\gamma_R)$. For the example of $|\alpha_{\rm targ} |=10\sqrt{2}$, which corresponds to a singlet fraction of $f_S=|\alpha_{\rm targ}|^2/(1+|\alpha_{\rm targ}|^2)=99.5\%$, we obtain a relaxation time of $\gamma\tau_{\rm rel}\simeq 300$, which is about two orders of magnitude longer than anticipated from the decay rate of the individual qubits.

    \subsubsection{Time-entanglement tradeoff}
    In the current setting, this behavior can be understood from  
    the following argument~\cite{Pichler_15}. After the initial quench, the system evolves quickly into a mixture of all possible two-qubit states, which are the dark state $|D\rangle$, the bright state $\ket{B} = (\alpha^*\ket{gg} + \ket{S})/\sqrt{1+|\alpha|^2}$, the doubly excited state $|ee\rangle$ and the triplet state $|T\rangle=(|eg\rangle+|ge\rangle)/\sqrt{2}$. Any further population of the dark state can only occur through the triplet state, which decays into state $|D\rangle$ with rate $\gamma|\langle D| \Sigma|T\rangle|^2$. In the limit $ |\alpha| \gg 1$, this rate is slow compared to the mixing of the states $|T\rangle$, $|B\rangle$ and $|ee\rangle$ by the driving field. Therefore, by assuming that the triplet state contains on average about $1/3$ of the residual population, we obtain the effective relaxation rate
    \begin{align}
        \gamma_{\text{rel}} =  \frac{2\gamma}{3(1+|\alpha|^2)},
        \label{eq: Relax_rate_Pichler}
    \end{align}
    at which the system approaches the steady state for long times. Clearly, for large drive amplitudes, the resulting relaxation time increases as $\tau_{\text{rel}}\sim|\alpha|^2$, diverging when one attempts to dissipatively prepare the maximally entangled singlet state (corresponding to $|\alpha|\rightarrow \infty$).

    Pocklington and Clerk~\cite{Pocklington_24} have shown that this trade-off between entanglement and preparation time is not a coincidence, but rather a consequence of a more general relation that holds for a broad class of  dissipative entanglement preparation schemes. Specifically, they defined the entanglement measure $\mathcal{E}_{2}$, which for a two-qubit system reduces to
    \begin{align}\label{eq: Scl_ent}
        \mathcal{E}_{2} =  2 \left(1 - \sum_{j=1}^2 p_j^2\right).
    \end{align}
    Here, $\sqrt{p_j}$ are the Schmidt coefficients of the steady state, such that $\mathcal{E}_{2}=0$  ($\mathcal{E}_{2}=1$) for a separable (maximally entangled) state. Under the assumption that the jump operators associated with the system's dissipation channels are sums of local operators,
    the rate of relaxation into this state is bounded from above by $\gamma_{\rm rel} \leq \Gamma_{\rm max}\sim (1-\mathcal{E}_{2})$, i.e., the rate vanishes for a near-maximally entangled steady state. In the current example, where $\mathcal{E}_{2} = |\alpha|^4/(1+|\alpha|^2)^2 = f_S^2$, we obtain 
    the upper bound (see Appendix \ref{app: T-E_bound}) 
    \begin{align}
    \gamma_{\rm rel} \leq \Gamma_{\rm max}\simeq \gamma(1-\mathcal{E}_2) \simeq \frac{2\gamma}{|\alpha|^2}
    \label{eq: Pow_law_Q}
    \end{align}
    in the limit of large $|\alpha|$. Up to a numerical prefactor, this upper bound follows the predicted trend for $\gamma_{\rm rel}$ in Eq.~\eqref{eq: Relax_rate_Pichler} and confirms the scaling $\tau_{\rm rel}\sim |\alpha|^2$ on more general grounds.
    
    In Fig.~\ref{fig: 2Qbit_Pichler}(c), we numerically verify this relation by evaluating the minimum time $T^{95\%}$ required to relax into the state $|D(\alpha_{\rm targ})\rangle$ with  a fidelity of at least $95\%$. This time is plotted against $1- \mathcal{E}_{2}(\alpha_{\rm targ})$ on a logarithmic scale, from which the inverse scaling expected from Eq.~\eqref{eq: Pow_law_Q} is immediately apparent. Based on the asymptotic exponential scaling in Eq.~\eqref{eq: Fid_exp_Pichler}, we also confirm that there is a near-perfect agreement with the analytic estimate for $\gamma_{\rm rel}$ in Eq.~\eqref{eq: Relax_rate_Pichler}. Note that in practice one is usually interested in preparing a state with a given singlet fraction $f_S$. To do so, the parameter $|\alpha_{\rm targ}|^2= cf_S/(1-f_S)$ must be set to a slightly higher value with $c>1$. For large $f_S\simeq 1$, the time to reach the condition $f_S = \langle S|\rho(T)|S\rangle$ is then approximately given by
    \begin{equation}
    \gamma T(f_S)\simeq  \frac{3c}{2(1-f_S)}\ln \left[ \frac{c}{(c-1)(1-f_S)}\right], 
    \end{equation}
    and contains additional prefactors and logarithmic corrections that further prolong the actual preparation time.

    \subsection{Entangled-state preparation with a linear sweep}
    The bound on the relaxation rate in Eq.~\eqref{eq: Pow_law_Q} has been derived under the assumption that---after the quench---the system evolves under a static Liouvillian, relaxing at a rate governed by its dissipative gap given in Eq.~\eqref{eq: Relax_rate_Pichler}. This raises the question whether or not the preparation time can be reduced by allowing for a more general time-dependence of the driving field $\Omega(t)$. A smooth ramp of $\Omega(t)\sim \sin^2(\pi t/(2T))$ has been considered in simulations in Ref.~\cite{Pichler_15}, but not with the explicit goal of achieving a speed-up in mind. In the following example, we consider an even simpler, linear ramp as shown in Fig.~\ref{fig: 2Qbit_Pichler}(a), with $\Omega^{\rm lin}(t)=\Omega_{\rm targ}t/T$.  
    
    In Fig.~\ref{fig: 2Qbit_Pichler}(b) we evaluate the state-preparation fidelity for this linear sweep for varying times $T$ and observe a significant improvement when compared to a quench. In Fig.~\ref{fig: 2Qbit_Pichler}(c) we also plot the corresponding preparation time $T^{95\%}$ as a function of $(1-\mathcal{E}_2)$ to study its dependence on the entanglement of the final state. 
    From these numerical simulations we find that for large $|\alpha_{\rm targ}|$, the preparation time follows closely the scaling 
    \begin{equation}
         T^{95\%}_{\rm lin}  \simeq \frac{\mathrm{C}}{\gamma \sqrt{1-\mathcal{E}_{2}}},
        \label{eq: Pow_law_lin}
    \end{equation}
    with a numerical prefactor of  $\mathrm C\simeq 4.09$, for $\mathcal{E}_2\rightarrow 1$. This improved scaling over the static case can be understood from the argument that the loss of fidelity should only depend on the rate of change, $\Omega_{\rm targ}/T$. As a consequence, for a fixed fidelity, we expect the scaling $T^{95\%}\sim \Omega_{\rm targ}$, while $(1-\mathcal{E}_2)\sim \Omega^{-2}_{\rm targ}$ for highly entangled states. Therefore, this simple linear sweep already demonstrates that time-dependent control fields can overcome the time-entanglement tradeoff of static preparation schemes. This observation also suggests that additional improvements can be obtained from more general time-dependent control pulses.

    \subsection{Leakage-minimizing paths}\label{subsec: LM_DD_ent_gen}
    
    We now apply the strategy of LM introduced in Sec.~\ref{sec: LM_for_Diss_sys} to the current setting. Using the fact that the steady states are pure, we can evaluate the instantaneous loss function defined in Eq.~\eqref{eq:loss_multi_param} and obtain (see Appendix~\ref{app: Qbit_like})
    \begin{align}
    \mathscr L(\Omega,\dot \Omega) =   \dfrac{2\dot{\Omega}^2(t)(\delta^2 + \gamma_A^2/4)}{\left(\delta^2 + 2\Omega^2(t)+  \gamma_A^2/4 \right)^2}. 
    \label{eq: loss_DD_entgen_2Q}
    \end{align}
    The corresponding Euler-Lagrange equation for $\Omega(t)$ reads
    \begin{align}\label{eq: EL_Pichler_2Q}
        \frac{\ddot{\Omega}(t)}{\dot{\Omega}(t)} = \frac{16\dot{\Omega}(t) \Omega(t)}{(4\delta^2 + 8\Omega^2(t) + \gamma_A^2)}
    \end{align}
    and determines the optimal $\Omega(t)$ that minimizes the leakage out of the prescribed adiabatic path, $\rho_{\rm ss}[\Omega(t)]$. For the boundary conditions $\Omega(0)=0$ and $\Omega(T)=\Omega_{\rm targ}$, the solution of Eq.~\eqref{eq: EL_Pichler_2Q} is given by (see Appendix~\ref{app:L_Invariance})
    \begin{equation}
        \Omega^{\rm LM}(t) = \frac{\sqrt{\delta^2 + \gamma_A^2/4}}{\sqrt{2}} \tan \left[ \arctan\left(|\alpha_{\rm targ}|\right)\frac{t}{T}  \right]
        \label{eq: Omega_t_2_qbit}
    \end{equation}
    and plotted as the red curve in Fig.~\ref{fig: 2Qbit_Pichler}(a).

    In Fig.~\ref{fig: 2Qbit_Pichler}(b) and (c), we compared the performance of the LM pulse with the quench and linear-ramp scenarios. As anticipated, the control pulse $\Omega^{\rm LM}(t)$ results in a significant improvement over the constant pulse and, in most cases, it also performs considerably better than a simple linear ramp. The difference between the three control schemes becomes particularly striking when considering the preparation of close to maximally entangled steady states, i.e., $(1-\mathcal{E}_2)\rightarrow 0$. In this regime, we find that the LM pulse achieves a preparation time $\gamma T^{95\%}\approx 30$ that is almost independent of the degree of entanglement of the targeted state. Therefore, LM overcomes even the weaker time-entanglement tradeoff observed for the linear ramp and accelerates the preparation process by orders of magnitude when compared to a simple quench.

    \subsection{Rapid preparation of multipartite entangled states}
    
    A key feature of LM is that in systems with an analytically known family of instantaneous steady states, the optimized control parameters can be derived for arbitrarily large quantum systems without knowledge of the spectrum of the Liouvillian or numerical solutions of the full master equation.  Indeed, for the current waveguide QED system, such an analytic solution for the steady state exists for an arbitrary even number of qubits $N$ coupled to a chiral or bidirectional waveguide, as long as certain detuning conditions are met. Therefore, the LM technique can be directly applied to speed up the preparation of such multiqubit entangled states.  
    
    \subsubsection{Dimerized and multipartite entangled states}
    Under the assumption of homogeneous driving, $\Omega_j=\Omega$, and that the detunings $\delta_j$ are pairwise anti-matched, i.e., for any qubit $j$ there is another qubit $k$ with $\delta_k=-\delta_j$, the master equation in Eq.~\eqref{eq: Pichler_ME} has a pure dark state of the form~\cite{Stannigel_12,Pichler_15}  
    \begin{equation}
    \begin{split}
        |\psi_D\rangle &= \left(\prod_p U_p\right) \bigotimes_{j=1}^{N/2}|D(\alpha_j)\rangle_{2j-1, 2j},
        \label{eq: Casc_Dim_Dark}
    \end{split}
    \end{equation}
    where $\ket{D(\alpha)}$ is a pairwise entangled state between neighboring qubits, as defined in Eq.~\eqref{eq: Dark_2qbit}. The state on the very right of Eq.~\eqref{eq: Casc_Dim_Dark} is a product of such pairwise entangled states with $\alpha_j = \sqrt{2}\Omega/(\delta_{2j-1}-i\gamma_A/2)$ determining the singlet component of each pair. As depicted in Fig.~\ref{fig: Pichler_Schematic}(b), this product is also the exact steady state for a pattern of alternating detunings $\delta_j$ for the qubits, which satisfy $\delta_{2j-1}=-\delta_{2j}$ for $0<j\leq N/2$. When other (pairwise anti-matched) detuning patterns are chosen [see Fig.~\ref{fig: Pichler_Schematic}(c)], the steady state is determined by additional unitaries of the form 
    \begin{equation}
        U_p = \exp{\Big(i\nu_p \vec{\sigma}_{j_p}\cdot\vec{\sigma}_{{j_p}+1}/2\Big)}.
        \label{eq: USWAP_Pichler}
    \end{equation}
    These unitaries implement partial swap operations between neighboring qubits. The rotation angles $\nu_p$ and the precise form of the resulting steady states depend on the specific detuning configuration, we refer to Refs.~\cite{Stannigel_12, Pichler_15} for a more detailed discussion. Importantly, in a chiral waveguide with $\gamma_A\neq 0$, one finds that $\nu_p\in (-\pi,\pi)$  and the total steady state is multipartite entangled when $|\delta_j|\neq 0$ and none of the nearest-neighbor detunings cancel pairwise. In a bidirectional scenario, where $\nu_p=0,\pm \pi/2$, this is not the case and the steady state remains bipartite entangled, although the entangled pairs can be redistributed across the chain.

    \subsubsection{Leakage-minimized paths for multipartite entangled states}
    \label{subsec: LM_DD_ent_gen_Nqbit}

    \begin{figure}[t]
        \includegraphics[width=\columnwidth]{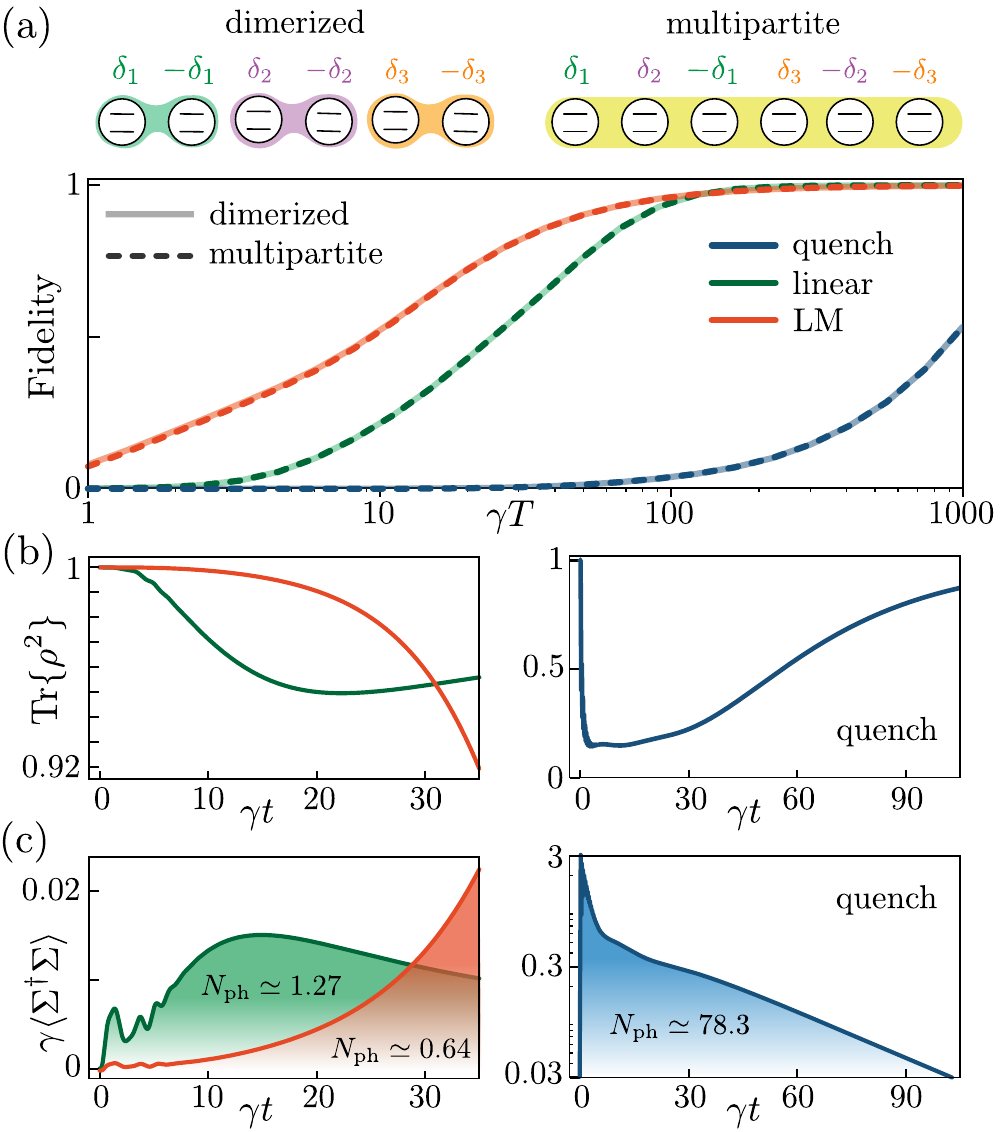}
        \caption{Rapid dissipative preparation of entangled multi-qubit states in a 1D waveguide. (a) Plot of the state-preparation fidelity $\mathcal{F}$ as a function of the final time $T$ and for $\Omega_{\rm targ}= 50.25\gamma$. The plot compares the achievable fidelities for the three different sweep functions for $\Omega(t)$ shown in Fig.~\ref{fig: 2Qbit_Pichler}(a) and for the two different detuning configurations indicated on top. These correspond to a dimerized steady state with two-qubit entanglement and to a more complex steady state with genuine multipartite entanglement, respectively. For the dimerized steady state, the evolution of the purity, ${\rm Tr}\{\rho^2\}$, and the normalized photon emission rate, $\langle \Sigma^\dag \Sigma\rangle$, are shown in (b) and (c). For these plots a lower value of $\Omega_{\rm targ}=10.05\gamma$ has been assumed to highlight the most important features. In all plots $(\delta_1, \delta_2, \delta_3)=(3\gamma, 4\gamma, 5\gamma)$ and in (c), the value of $N_{\rm ph}$ indicates the total number of emitted photons, as defined in Eq.~\eqref{eq:Nph}.} 
        \label{fig: 6qbit_fid_comp}
    \end{figure}
    
    A convenient property of the states in Eq.~\eqref{eq: Casc_Dim_Dark} is that the unitaries $U_p$ do not depend on the driving strength. This implies that for a control scheme that only involves $\Omega(t)$, we can use the fact that $\partial_t U_p=0$ to show that the $U_p$ cancel in the evaluation of the loss function in Eq.~\eqref{eq:loss_multi_param}. As a consequence, the loss function is independent of the entanglement structure of the targeted steady state and we obtain (see Appendix~\ref{app: Qbit_like})
    \begin{equation}
    \mathscr L(\Omega,\dot \Omega) = \sum_{j=1}^{N} \dfrac{\dot{\Omega}^2(t)\big(\delta_j^2 + \gamma_A^2/4\big)}{\left[\delta_j^2 + \gamma_A^2/4 + 2\Omega^2(t)\right]^2},
    \end{equation}
    where pairwise anti-matched detunings have been assumed. The solution of the corresponding Euler-Lagrange equation can be written in the form (see Appendix~\ref{app:L_Invariance})
    \begin{equation}
    \begin{split}
       \Omega^{\rm LM}(t)= \mathcal{I}^{-1}\left(  \mathcal{I}(\Omega_{\rm targ})\frac{t}{T} \right),
        \label{eq:Omega_multi}
    \end{split}    
    \end{equation}
    where, for $\Omega(0)=0$, 
    \begin{equation}
       \mathcal{I}(\Omega) =  \int_{0}^{\Omega} d\tilde \Omega \sqrt{\sum_{j=1}^{N} \frac{\delta_j^2 + \gamma_A^2/4}{(\delta_j^2+ \gamma_A^2/4 + 2\tilde\Omega^2)^2}}.
    \end{equation}
    When $\delta_j=\pm \delta$, this integral can be solved exactly and we recover the optimal driving pulse $\Omega^{\rm LM}(t)$ given in Eq.~\eqref{eq: Omega_t_2_qbit} above. In general, $\mathcal{I}(\Omega)$ can be solved and inverted numerically to obtain $\Omega^{\rm LM}(t)$ in Eq.~\eqref{eq:Omega_multi}. Notably, since this equation only depends on the rescaled time $t/T$, it is enough to solve it once to obtain the optimized pulse for arbitrary $T$. Furthermore, the formal solution can be obtained for arbitrary $\delta_j$ and $\gamma_A$ through one numerical integral evaluation. This is in stark contrast to conventional optimal control schemes, which require a separate optimization cycle for each value of $T$, $\delta_j$, and $\gamma_A$.
    
    In Fig.~\ref{fig: 6qbit_fid_comp} we illustrate the application of LM for the preparation of bi- and multipartite entangled states for a chain of $N=6$ qubits. In these plots, we consider a unidirectional waveguide with $\gamma_R=\gamma$ and $\gamma_L=0$ and two different detuning configurations, shown at the top of Fig.~\ref{fig: 6qbit_fid_comp}. They correspond to a steady state with pairwise entanglement between nearest neighbors and a genuine multipartite entangled steady state, respectively. In Fig.~\ref{fig: 6qbit_fid_comp}(a) we plot the fidelity $\mathcal{F}$ as a function of the preparation time $T$, comparing quench, linear and LM pulses. As in the two-qubit case, we observe a similar speedup for the preparation of entangled multiqubit states when applying linear sweeps or LM pulses. This speedup and the resulting fidelities are independent of the detuning pattern and therefore the same optimized control pulses can be used to distribute bi- or multipartite entanglement across an arbitrary number of qubits. This feature can be traced back to the form-invariance~\cite{Stannigel_12} of the waveguide master equation under the unitary transformations in Eq.~\eqref{eq: USWAP_Pichler}, and is explicitly verified also for time-dependent control fields by our numerical simulations.

    \subsubsection{Relaxation vs adiabatic state preparation} 
    Apart from the final state preparation fidelity, it is instructive to also compare the evolution of the system toward the targeted state~\cite{Pichler_15}. This is done in Fig.~\ref{fig: 6qbit_fid_comp}(b) and (c), where we plot the purity, ${\rm Tr}\{\rho^2(t)\}$, and the expectation value $\langle \Sigma^\dag \Sigma\rangle(t)$, where $\Sigma$ is the collective jump operator in Eq.~\eqref{eq: Pichler_ME}, as a function of time and for the three different sweep functions. Here, $\langle \Sigma^\dag \Sigma\rangle(t)$ corresponds to the normalized rate of photon emission into the waveguide, and we define by
    \begin{equation}\label{eq:Nph}
    N_{\rm ph}=\gamma \int_0^T  dt \, \langle \Sigma^\dag \Sigma\rangle(t)
    \end{equation} 
    the total number of emitted photons during the preparation time.
    
    From the various plots, we see, first of all, a striking difference between the dynamics after the quench and for the time-dependent sweeps~\cite{Pichler_15}. In the first case, the system is quickly driven into a highly mixed state, after which it relaxes into the dark state through the emission of many photons. In contrast, when applying the LM pulse, the system remains very close to the family of pure dark states and almost no photons are emitted. Thus, it follows the instantaneous steady state adiabatically, with only small non-adiabatic corrections at the final stage of the evolution.  
    
    The linear sweep produces dynamics that lie between the other two cases. Although the system remains close to a pure state throughout the evolution, stronger nonadiabatic corrections arise at early times. These excitations are subsequently reduced---at least in part---by relaxation during the later stages of the protocol. This example also highlights an important limitation of LM. The method minimizes transitions away from the manifold of instantaneous steady states, but does not account for relaxation processes that drive the system back to this manifold. As a result, LM is not necessarily optimal, especially for less entangled target steady states which permit faster relaxation. In particular, for sufficiently long evolution times, $\gamma_{\rm rel}T\gtrsim 1$, a protocol that induces a larger initial disturbance can outperform adiabatic state preparation when the ensuing relaxation time is long enough. Note that the relation between adiabatic transfer and relaxation processes has also been analyzed in Ref.~\cite{Venuti_17} for various model systems with different conclusions. In the current setup, when $\Omega_{\rm targ}$ is large and for short preparation times, $T\ll \gamma_{\rm rel}^{-1}$, LM provides a particularly effective control strategy, as relaxation back to the family of target states is slow and  constitutes the main bottleneck in the dynamics.  
    


    \section{Rapid entanglement distribution through a nonlocal squeezed reservoir}
    \label{sec: TMS_ent_gen}
    \begin{figure}
        \centering
     \includegraphics[width=\linewidth]{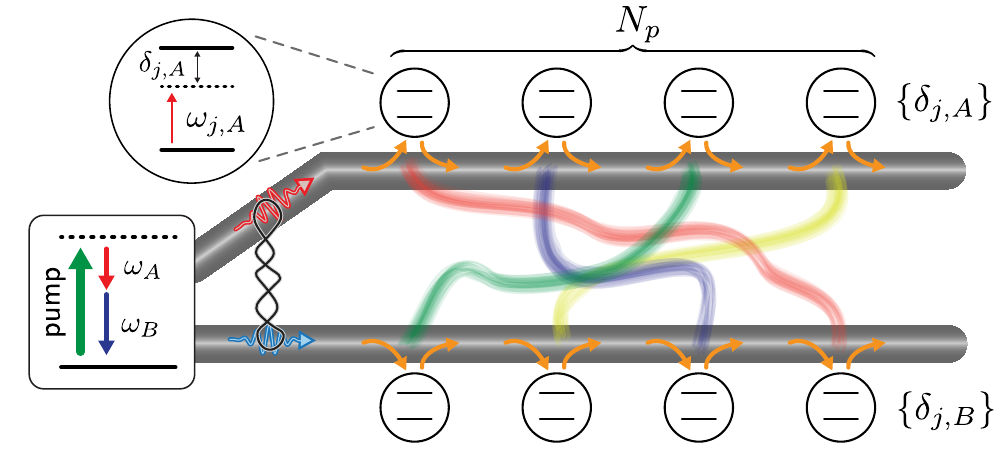}
        \caption{Schematic of the dual-waveguide setup considered in Sec.~\ref{sec: TMS_ent_gen}. Here, a broadband parametric amplifier (box on the left) produces a TMS photonic reservoir, which relaxes the qubits coupled to the waveguides into  steady states with entanglement distributed across the two subsystems. Different types of bi- and multipartite entanglement patterns can be generated by appropriately chosen detunings $\delta_{j,A}$ and $\delta_{j,B}$.  See the text and Ref.~\cite{Agusti_23} for more details.}
        \label{fig:Sch_Nqbit_TMS}
    \end{figure}
    
    As a second important example for the application of LM, we consider the preparation of remote entangled qubit states through a TMS reservoir~\cite{Kraus_04}. Here, instead of exchanging photons along a single waveguide, the qubits are immersed in a distributed photonic bath with preexisting squeezing correlations, as illustrated in  Fig.~\ref{fig:Sch_Nqbit_TMS}. Such a reservoir can be generated, for example, by the output of a nondegenerate parametric amplifier~\cite{Couteau2018,Eichler2011,Flurin2012,Menzel2012,Juanes_26}. 
    As long as the bandwidth of the amplifier is sufficiently large \cite{Agusti_22}, the quantum correlations of the photons are mapped onto the qubit states and---under otherwise ideal conditions---the resulting steady state of the qubits is pure and can be either bipartite or multipartite entangled~\cite{Agusti_23}. An experimental implementation of this scheme was recently demonstrated, using squeezed microwave photons to entangle two separated superconducting qubits~\cite{Juanes_26}.

    \subsection{Qubits coupled to a squeezed reservoir}
    
    For the following discussion, we consider a setup as depicted in Fig.~\ref{fig:Sch_Nqbit_TMS}, where a total number of $N$ qubits, evenly split into $N_p=N/2$ pairs, are coupled to two independent waveguides labeled $A$ and $B$. For concreteness, in this section we focus exclusively on the case of unidirectional waveguides with $\gamma_R=\gamma$ and $\gamma_L=0$. Each waveguide is driven by one of the two outputs of a nondegenerate parametric amplifier to prepare a broadband TMS reservoir. After adiabatically eliminating the photonic degrees of freedom, the remaining dynamics of the qubits is described by the master equation~\cite{Agusti_23}
    \begin{equation}
    \begin{split}
            \dot{\rho} =& -i \sum_{\substack{j=1 \\ \eta = A,B}}^{N_p} \left[\frac{\delta_{j, \eta}}{2} \sigma^z_{j,\eta} - \frac{i\gamma}{2}\sum_{k>j}\Big( \sigma^+_{k, \eta} \sigma^-_{j, \eta} -  {\rm h.c} \Big), \rho \right] \\
            &+ \gamma \sum_{\eta = A,B} \mathcal{D}[J_\eta]\rho   + \sum_{\substack{j=1 \\ \eta = A,B}}^{N_p}  \gamma_\phi \mathcal{D}[\sigma^z_{j,\eta}]\rho \equiv \mathcal{L}\rho.
            \label{eq: ME_TMS_NQbit}
    \end{split}        
    \end{equation}
     Here, the first line accounts for the detunings and the cascaded interactions within each waveguide. In the second line of Eq.~\eqref{eq: ME_TMS_NQbit}, we have introduced the collective jump operators
    \begin{align}
        J_A &= \cosh(r) \Sigma_A -  \sinh(r)\Sigma_{B}^\dag,\\
         J_B & = \cosh(r) \Sigma_B -  \sinh(r)\Sigma_{A}^\dag,
    \end{align}
    where $\Sigma_\eta  = \sum_{j=1}^{N_p} \sigma^-_{j,\eta}$. They reflect the collective qubit-photon interactions along each channel and the squeezing correlations between the two waveguides. The squeezing strength $r$ is controlled, for example, by the pumping strength of the parametric amplifier. Finally, the last term in Eq.~\eqref{eq: ME_TMS_NQbit} accounts for a dephasing of all qubits with rate $\gamma_\phi$, which we will use below to model a more realistic experimental scenario. For now, we set $\gamma_\phi=0$.

    \subsection{Distribution of two- and multi-qubit entanglement}
    In the simplest case, $N_p=1$, where each waveguide only contains a single qubit, and assuming $\delta_{1,A} = -\delta_{1,B}$, the unique steady state of the master equation in Eq.~\eqref{eq: ME_TMS_NQbit} is the pure entangled state~\cite{Kraus_04}
    \begin{align}
     |\psi_D\rangle = |\phi^+(r)\rangle = \frac{\cosh(r)|gg\rangle + \sinh(r)|ee\rangle}{\sqrt{\cosh(2r)}},
     \label{eq: Dark_2Qbit_TMS}
    \end{align}
    which approaches a maximally entangled Bell state when $r\gg1$.
    For this state, the scaled entanglement defined in Eq.~\eqref{eq: Scl_ent} evaluates to
    \begin{align}
        \mathcal{E}_{2} = \tanh^2(2r).
        \label{eq: scl_ent_TMS}
    \end{align}
    The asymptotic relaxation rate into the steady state, $\gamma_{\rm rel}\simeq \lambda_{\rm dis}$, can be estimated from the dissipative gap $\lambda_{\rm dis}= -\max_{\ell_a\neq0}\operatorname{Re}\{ \ell_a\}$,
    where $\ell_a$ are the eigenvalues of the Liouvillian $\mathcal{L}$. For $N_p=1$ and $r\gtrsim 2$, we obtain 
    \begin{align}
    \gamma_{\text{rel}}(r) \simeq \frac{2\gamma}{3} \sqrt{1-\mathcal{E}_{2}(r)},
        \label{eq: T_E_TMS}
    \end{align}
    which seemingly violates the time-entanglement tradeoff for static Liouvillians discussed above. As shown in Appendix~\ref{app: T-E_bound}, this result can be understood---in agreement with Ref.~\cite{Pocklington_24}---from the fact that not only the degree of entanglement, but also the form of the jump operators varies with $r$. However, despite exhibiting a different scaling, a tradeoff between the preparation time and the entanglement of the final steady state also holds in this setting. In particular, Eq.~\eqref{eq: T_E_TMS} implies the scaling $\tau_{\rm rel}\sim e^{2r}$ for a static TMS reservoir. 
    
    Similar to the case of a single waveguide, for pairwise anti-matched detunings with $\delta_{j,A}=-\delta_{k,B}$, the squeezed dual-waveguide 
    system supports a family of pure bi- and multipartite entangled states, which  are of the form~\cite{Agusti_23}
    \begin{equation}\label{eq:Dark_NQbit_TMS}
    \begin{split}
        |\psi_D\rangle &= \left(\prod_p U_p\right) \bigotimes_{j=1}^{N_p}|\phi^+(r)\rangle_{j, j}.
    \end{split}
    \end{equation}
    Here, $|\phi^+(r)\rangle_{j, j}$ denotes a pairwise entangled state as in Eq.~\eqref{eq: Dark_2Qbit_TMS} between the $j$-th qubits in waveguide A and waveguide B. The $U_p$ are partial swap operations, as defined in Eq.~\eqref{eq: USWAP_Pichler}, which act on neighboring qubits in the same waveguide. For $\delta_{j,A}=\delta_{j,B}=0$, the $U_p=\mathbbm{1}$ and $|\psi_D\rangle$ represents a product of Bell states across the two waveguides. For $|\delta_{j,A}|\sim \gamma$ and by matching the detunings of qubits in waveguide A with different qubits in waveguide B, the $U_p$ act as nontrivial entangling operations and in general $|\psi_D\rangle$ is a complex multipartite entangled state.
    
    In Ref.~\cite{Agusti_23} it was shown that for $\delta_{j,A}=0$, the entangled pairs form sequentially, such that the total preparation time scales approximately as $\tau_{\rm rel}\sim N_p/\gamma_{\rm rel}(r)$. For very far detuned qubits with $|\delta_{j,A}-\delta_{k,A}|\gg \gamma$ for $j\neq k$, the different matched pairs do not affect each other and $\tau_{\rm rel}$ becomes approximately independent of $N_p$. However, these conditions do not lead to multipartite entanglement, and in realistic settings, the maximally allowed qubit detuning is also limited by the bandwidth of the TMS source. Therefore, for many cases of interest, the scaling $\tau_{\rm rel}\sim N_p e^{2r}$ is expected.

    \subsection{Leakage minimization for squeezed reservoirs}
    \begin{figure}[t]
        \centering
        \includegraphics[width=\columnwidth]{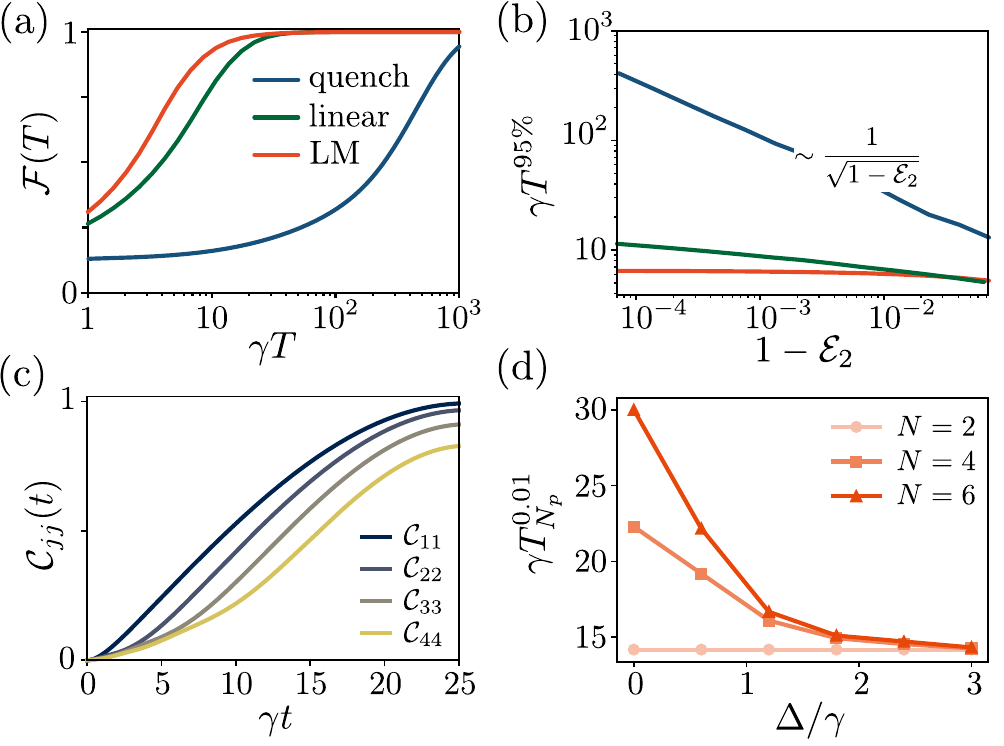}
        \caption{Dissipative preparation of entangled multi-qubit states in a TMS reservoir. In all plots, a set of linearly increasing, pairwise anti-matched detunings $\delta_{j,A} =-\delta_{j,B} = (j-1)\Delta$ is assumed, such that the resulting steady state is a product of Bell states, $|\psi_D\rangle=\bigotimes_j |\phi^+(r_{\rm targ})\rangle_{j,j}$.   
        (a) The fidelity $\mathcal{F}(T)$ for preparing $N_p=3$ Bell pairs is plotted for $\Delta/\gamma=2$ as a function of the total protocol time $T$ and for three different control pulses for the squeezing parameter $r(t) \in [0,r_{\rm targ}=3]$. (b) For a single pair of qubits, $N_p=1$, the time $T^{\rm 95\%}$ to reach a fidelity of $\mathcal{F}\geq 0.95$ is plotted as a function of the entanglement $\mathcal{E}_2$ of the final steady state for $r_{\rm targ}\in[1,3]$. 
        (c) Time evolution of the concurrences $\mathcal{C}_{jj}=\mathcal{C}(\rho_{j,j}^{AB})$ of the reduced two-qubit states $\rho_{j,j}^{AB}$ for $\Delta=0$ and $r_{\rm targ}=2$. (d) Plot of the preparation time $\bar T_{N_p}^{0.01}$ for different numbers of entangled pairs and varying $\Delta$. Here, $\bar T_{N_p}^{0.01}$ defines the time at which the infidelity per Bell pair, $\zeta/N_p$, first reaches a value below $0.01$. 
        }
        \label{fig: Combined_plots_TMS}
    \end{figure}
    To overcome this unfavorable scaling for a static TMS reservoir, we apply again the formalism of LM to identify an optimal time-dependent shape for the squeezing parameter $r(t)$. Note that in contrast to the previous examples in Sec.~\ref{sec: DD_ent_gen}, this variation corresponds to a control of the dissipators, rather than the Hamiltonian. 
    Since the squeezing dissipators are uniform along the waveguide and also $\partial_r U_p=0$, we can apply the LM formalism directly to the family of many-body steady states in Eq.~\eqref{eq:Dark_NQbit_TMS} and obtain 
    \begin{align}
        \mathscr{L}(r, \dot{r})= N_p \dot{r}^2 \sech^2(2r).
        \label{eq: loss_TMS_NQbit}
    \end{align}
    For an initial squeezing strength $r_0$ and a targeted squeezing strength $r_{\rm targ}$, the solution of the corresponding Euler-Lagrange equation is (see Appendix~\ref{app:L_Invariance})
    \begin{align}
    r^{\text{LM}}&(t) = \, \frac{1}{2}\ln\left(\tan\left\{\arctan(e^{2r_0}) \phantom{\frac{t}{T}}\right.\right.\label{eq: r_LM_TMS} \\
    \phantom{  r^{\text{LM}}} & \left. \left.+ \left[\arctan\left(e^{2r_{\mathrm{targ}}}\right)- \arctan\left(e^{2r_0}\right)\right]\frac{t}{T}\right\}\right),\nonumber
    \end{align}
    which is independent of $N$ and the choice of detunings. This property follows from the product structure of the dark state in Eq.~\eqref{eq:Dark_NQbit_TMS} and the fact that the Bell state in Eq.~\eqref{eq: Dark_2Qbit_TMS} depends on the squeezing parameter only. 
    
    In Fig.~\ref{fig: Combined_plots_TMS}(a) we simulate the state preparation fidelity $\mathcal{F}$ for a pair-wise entangled steady state with $N_p=3$ qubits per waveguide and $r_{\rm targ}=3$. This plot compares again three different control schemes for the squeezing strength $r(t)$, which correspond to a quench, a linear ramp, and the LM pulse, respectively. Also, in this example we find that a time-dependent ramp of the squeezing strength can reduce the preparation time by more than two orders of magnitude for highly entangled states, with LM outperforming a simple linear increase. 
    
    For the case of a single pair, $N_p=1$, we plot [Fig.~\ref{fig: Combined_plots_TMS}(b)] the preparation time $T^{95\%}$ as a function of the entanglement of the targeted steady state. This plot verifies the predicted scaling in Eq.~\eqref{eq: T_E_TMS} for a static TMS reservoir and shows a saturation of the preparation time when the LM pulse is used.    
    
    As mentioned above, a characteristic feature of a unidirectional TMS reservoir is that for $\delta_{j,\eta}=0$, the Bell pairs are purified sequentially. In turn, for far detuned pairs, the purification happens in parallel, employing squeezing correlations of the reservoir at different frequencies. The plots in Fig.~\ref{fig: Combined_plots_TMS}(c) and (d) demonstrate that this feature persists under LM control, but on much shorter timescales. As shown in Fig.~\ref{fig: Combined_plots_TMS}(c) for resonant qubits, the individual Bell pairs are purified sequentially, but also the amount of entanglement increases gradually as the squeezing strength is increased. Note that for this plot, the total time $T$ is chosen deliberately short to highlight the fact that Bell pairs further away from the sources will be affected more strongly by nonadiabatic corrections. When the detuning between the individual pairs is increased, the state preparation becomes essentially independent of the number of entangled pairs, similar to the static case [see Fig.~\ref{fig: Combined_plots_TMS}(d)]. This demonstrates that even though the optimized LM pulse is the same for all detunings, the achieved speedup is fully compatible with the intrinsic parallelizability of entanglement-distribution in TMS reservoirs.

    \subsection{State preparation versus dephasing}
    \label{subsec: TMS_Deph}
    
    So far---both in Sec.~\ref{sec: DD_ent_gen} and in the current setting---we have studied the dissipative preparation of entangled states under ideal conditions, where even in the static case the targeted steady state is reached eventually. This is no longer the case in a more realistic scenario, where the engineered dissipation process competes with other unwanted decay channels or dephasing processes. To illustrate the influence of such unwanted decoherence processes, we return to the case of two qubits coupled to a TMS reservoir and include the effect of Markovian dephasing with rate $\gamma_\phi>0$ in Eq.~\eqref{eq: ME_TMS_NQbit}. Note that dephasing reduces the coherence between the states $|gg\rangle$ and $|ee\rangle$ and is therefore in direct competition with the engineered TMS reservoir, which stabilizes a particular superposition of those states. We expect that the dissipative stabilization of entanglement prevails as long as $\gamma_\phi \ll \tau_{\rm rel}^{-1}$, which already indicates the stringent requirement  that static state-preparation schemes impose on $\gamma_\phi$.
    
    \begin{figure}[t]
        \centering   
        \includegraphics[width=\columnwidth]{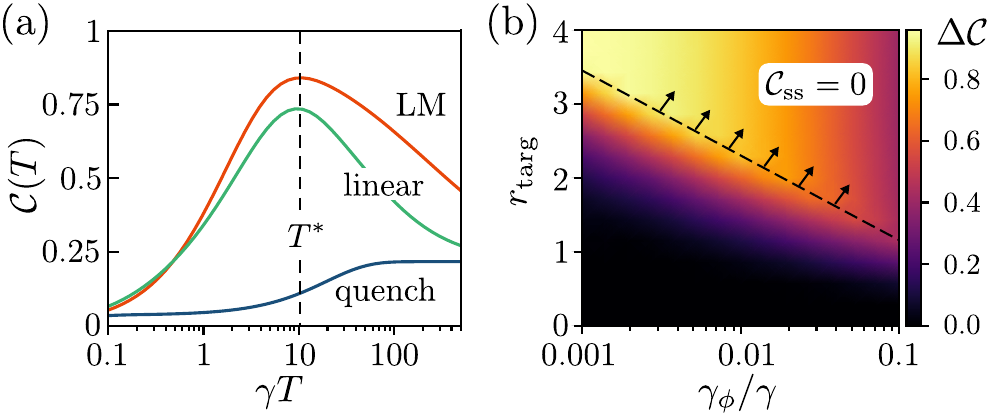}
        \caption{Dissipative preparation of an entangled two-qubit state in a TMS reservoir and in the presence of dephasing.  (a) The concurrence $\mathcal{C}$ of the final state $\rho(T)$ is evaluated as a function of the preparation time $T$ for $r_{\rm targ}=2$ and a dephasing rate $\gamma_\phi/\gamma=0.01$. The plot compares the quench dynamics with a linear ramp of $r(t)$ and the LM pulse $r^{\rm LM}(t)$. The dashed vertical line indicates the optimal time $T^*$ obtained for LM. (b) The difference $\Delta \mathcal{C}=\mathcal{C}_{\rm op}-\mathcal{C}_{\rm ss}$ between the optimal concurrence $\mathcal{C}_{\rm op}=\mathcal{C}(T^*)$ reached with LM and the steady concurrence is shown for different values of $r_{\rm targ}$ and $\gamma_\phi$. The dashed line marks the boundary above which the steady-state entanglement vanishes. }
        \label{fig: C_max_TMS}
    \end{figure}
    
    In Fig.~\ref{fig: C_max_TMS}(a) we plot the concurrence of the two-qubit state after a given preparation time $T$, assuming $r_{\rm targ}=2$ and a small but nonvanishing dephasing rate $\gamma_\phi/\gamma=0.01$. As above, we compare the static case with a linear ramp of $r(t)$ and the optimized pulse $r^{\rm LM}(t)$. In all cases, dephasing reduces the maximal amount of entanglement, which in the static case is reached under most conditions asymptotically for very long times. Note that despite a rather small dephasing rate, this steady state entanglement is reduced to a value of $\mathcal{C}_{\rm ss}\approx 0.22$, compared to $\mathcal{C}_{\rm ss}\approx 0.999$ in the ideal case. For the time-dependent control parameters, the concurrence depends non-monotonically on the preparation time, with a maximal value reached at an intermediate optimal preparation time $T^*$. While still being degraded by dephasing, the maximally achievable entanglement is considerably more robust than in the static case, with LM control outperforming the linear ramp under most conditions. Furthermore, the maximal concurrence achievable with either time-dependent control scheme satisfies $\max_T\mathcal{C}(T)\geq\mathcal{C}_{\rm ss}\simeq\mathcal{C}(T\gg\gamma_\phi^{-1})$.
    
    In Fig.~\ref{fig: C_max_TMS}(b) we evaluate the optimal concurrence, $\mathcal{C}_{\rm op} ={\rm max}_{T}\{ \mathcal{C}(T) \} =\mathcal{C}(T^*)$, as obtained from LM for different squeezing parameters $r_{\rm targ}$ and dephasing rates $\gamma_\phi$. This value is compared to the steady-state concurrence $\mathcal{C}_{\rm ss}$, which can be reached using noisy qubits coupled to a static TMS reservoir. This plot of the difference $\Delta \mathcal{C} =\mathcal{C}_{\rm op}-\mathcal{C}_{\rm ss}$ shows the actual benefit of LM over static entanglement preparation schemes, which in a large region of the considered parameter regime (above the dashed line) would not generate entanglement at all. Qualitatively similar conclusions also hold for $N_p>1$ and for the single-waveguide setup discussed in Sec.~\ref{sec: DD_ent_gen}.

    \section{Optimality and counterdiabatic control}
    \label{sec: QOC_CD}
    
    While for the examples considered above, the LM approach provides a substantial speedup over static relaxation into the steady state, the control pulses $\boldsymbol{\lambda}^{\rm LM}(t)$ obtained in this way are not guaranteed to be optimal. Indeed, we have already seen that a simple linear sweep can outperform LM under certain conditions. Therefore, in this section we benchmark LM against conventional optimal control techniques and also investigate the potential benefit of applying additional CD control.

    \subsection{Comparisons with optimal control}
    \label{subsec: QOC_DD_ent_gen}
    For small systems, the task of finding control parameters for preparing a targeted state $\rho_{\rm targ}$ with the highest fidelity can be formulated as a general optimal control problem with a cost function 
    \begin{equation}
    \zeta = 1- \mathcal{F}\left(\rho(T)|\rho_{\rm targ}\right)
    \end{equation}
    that compares the time-evolved state with the targeted one. There are many standardized methods that have been developed to solve this and similar state preparation problems (see, for instance, Refs.~\cite{Koch_22, Gautier_25, Petruhanov_2023, Caneva_11, Machnes_18}). For the following comparison with LM, we focus on the \ac{GOAT} method introduced in Ref.~\cite{Machnes_18}, but we do not expect any of the main conclusions to depend on this specific choice. More details about the GOAT method adapted to open quantum systems are provided in Appendix~\ref{app: GOAT}. 
    
\begin{figure}
        \centering
        \includegraphics[width=\columnwidth]{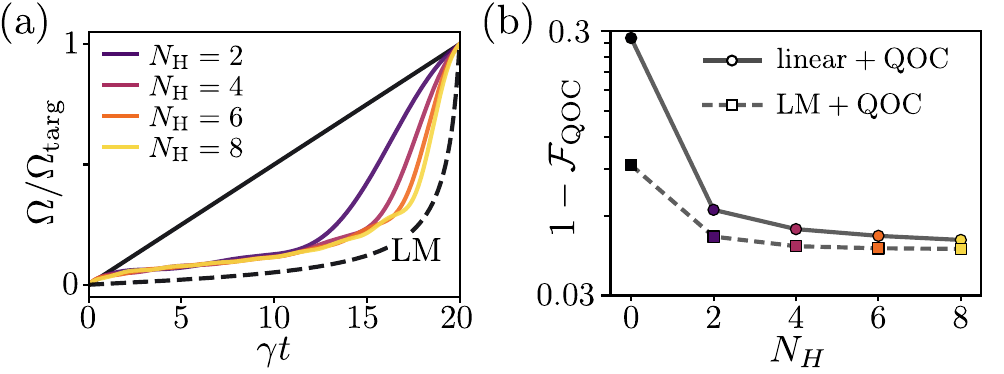}
        \caption{Quantum optimal control. (a) Plot of the optimized pulse shape $\Omega(t)$ obtained from QOC for different numbers of harmonics $N_H$ in the expansion in Eq.~\eqref{eq: QOC_ansatz} and using a linear ramp (black solid line) as a seed. The dashed line indicates the shape of the LM pulse, $\Omega^{\rm LM}(t)$. (b) The final state preparation fidelity $\mathcal{F}_{\rm QOC}$ is plotted as a function of the number of harmonics, using either the linear sweep (solid line) or the LM pulse (dashed line) as a seed. For both plots, we have considered a system of $N=4$ qubits coupled to a single unidirectional waveguide with $\gamma T=20$, $\Omega_{\rm targ}/\gamma = 50.25$ and detunings $(\delta_1,\delta_2,\delta_3,\delta_4)/\gamma= (3,-3,5,-5)$.}
        \label{fig: QOC_Stannigel}
    \end{figure}

    As a concrete example, we consider the case of $N=4$ qubits coupled to a single waveguide with nonvanishing detunings as specified in the caption of Fig.~\ref{fig: QOC_Stannigel}. As above, we assume a single control parameter $\Omega(t)$, which we expand as a finite Fourier series,
    \begin{equation}
    \begin{split}
        \Omega(t) =\, \Omega_{\rm seed}(t) 
        + \sum_{n=1}^{N_H} \Big(& a_n \sin\left(\omega_n t)\right) \\
        &+ b_n \left[\cos\left(\omega_n  t\right)-1\right]\Big),
        \label{eq: QOC_ansatz}
    \end{split}
    \end{equation}
    where $\omega_n=2\pi n /T$. 
    Here, $\Omega_{\rm seed}(t)$, with $\Omega_{\rm seed}(0)=0$ and $\Omega_{\rm seed}(T)=\Omega_{\rm targ}$, is used as an initial seed function for the optimizer, which then searches for optimal values for the remaining set of $2N_H$ Fourier coefficients $\{a_n, b_n\}$.

    In Fig.~\ref{fig: QOC_Stannigel}(a), we use QOC to find the optimal control pulse $\Omega(t)$ for preparing a pairwise entangled state of $N=4$ qubits. Using a linear ramp as a seed, we see that as more and more harmonics are added to the expansion in Eq.~\eqref{eq: QOC_ansatz}, the optimized pulse shape starts to 
    deviate significantly from the linear sweep and becomes close to---but not identical to---the LM pulse. In particular, the optimal pulse still shows a sharp increase at the very end of the protocol, but exhibits an additional small bump  at the beginning of the pulse. While this initial step induces nonadiabatic excitations, the system still has enough time to relax back into the pure steady state.  For the same example, we evaluate in Fig.~\ref{fig: QOC_Stannigel}(b) the preparation fidelity $\mathcal{F}_{\rm QOC}(T)$, as obtained from QOC for different numbers of harmonics. In this plot, we compare the case from above, where $\Omega_{\rm seed}(t)=t \Omega_{\rm targ}/T$, with the case where we use the LM pulse as a seed, i.e., $\Omega_{\rm seed}(t) =\Omega^{\rm LM}(t)$. While for both seed functions the optimized fidelity converges to the same value of $\mathcal{F}_{\rm QOC}(T)\simeq 0.965$ for large enough $N_H$, this value is approached much faster when seeded with the LM pulse.

    In summary, this concrete example demonstrates that the LM approach delivers a control pulse that is already close to the optimal solution found with fully numerical pulse-optimization tools. We emphasize though that such numerical search algorithms require the integration of the full master equation many times, and such methods are no longer applicable for slightly larger qubit numbers. 
    Even in cases where optimal control can be applied, the LM path provides an informed initial guess, which improves the performance of the successive numerical optimization step. In Appendix~\ref{app: GOAT} we discuss a second example, which shows that similar conclusions about the optimality of LM are also obtained for the tunable TMS reservoir.

    \subsection{Counterdiabatic driving}
    \label{subsec:CD driving}
    In Sec.~\ref{sec: LM_for_Diss_sys} we have shown that LM can be generalized to account for external CD control fields as an additional control knob to suppress nonadiabatic transitions of the states out of their predefined path. Here we briefly illustrate the application of this approach for  a chain of $N=6$ qubits coupled to a single bidirectional waveguide with $\gamma_L = \gamma_R$. We consider anti-matched detunings (such that for every index $i$, there is a unique index $j$ such that $\delta_i+\delta_j=0$) between qubits such that the steady state is a product of entangled dimers, as shown in Fig.~\ref{fig: Pichler_Schematic}(b). In addition to the master equation specified in Eq.~\eqref{eq: Pichler_ME}, we include CD driving fields with Hamiltonian
    \begin{align} \label{eq: local_CD_sigma_y}
        H_{\rm CD}(t) = \sum_{j=1}^{N} \beta_j(t)\sigma^y_{j}.
    \end{align}
    As shown in Appendix~\ref{app: CD_Dimer}, for the current example, this is the only possible local Hamiltonian that reduces the loss function. While in principle additional two-qubit interactions could be considered~\cite{Lim_24}, such control terms are usually not available in entanglement distribution schemes for spatially separated qubits.
    
    The addition of these controls modifies the form of the loss function $\mathscr{L}(\Omega, \dot{\Omega}, \{\beta_j\})$ according to Eq.~\eqref{eq:LossFunction_CD_Pure}. For the current example, the derivation of the resulting expression of $\mathscr{L}$ is outlined in Appendix~\ref{app: CD_Dimer}. Importantly, it does not depend on the velocities $\dot \beta_j$, such that the optimal control parameters are determined by $\partial \mathscr{L}/\partial \beta_j=0$, yielding 
    \begin{align}
    \beta_j(t) = \frac{-\dot{\Omega}(t)\delta_j}{\delta_{j}^2 + 4\Omega^2(t)}.
    \label{eq: beta_j_CD}
    \end{align} 
    After substituting this result back into $\mathscr{L}(\Omega, \dot{\Omega}, \{\beta_j\})$, we obtain an effective loss function $\mathscr{L}_{\rm eff}(\Omega, \dot{\Omega})$ for the driving strength $\Omega(t)$ only. Its full expression is given in Eq.~\eqref{eq: Loss_eff_CD}. From the solution of the corresponding Euler-Lagrange equations, we then obtain $\Omega^{\rm LM}(t)$ and the CD controls using  Eq.~\eqref{eq: beta_j_CD}.  
    
    \begin{figure}[t]
        \centering
        \includegraphics[width=1.01\linewidth]{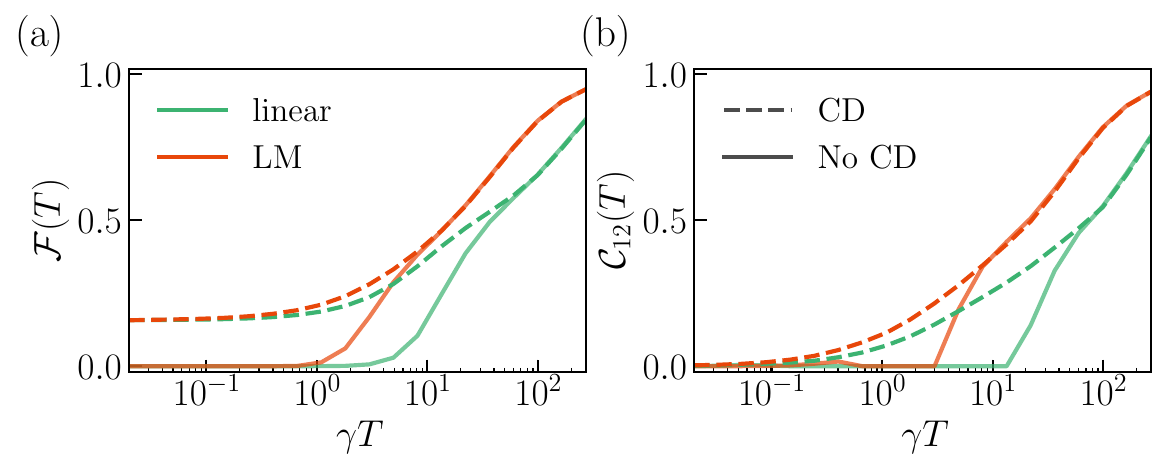}
        \caption{Plot of (a) the state preparation fidelity and (b) the concurrences of the reduced state of qubits $j=1$ and $j=2$ as a function of the total sweep time $T$. For both plots, we have considered a set of $N=6$ qubits coupled to a single bi-directional waveguide with $\gamma_R=\gamma_L=\gamma/2$, $\Omega_{\rm targ}/\gamma = 25$ and detunings $(\delta_1,\delta_2,\delta_3,\delta_4,\delta_5,\delta_6)/\gamma= (1.5,-1.5,2,-2,2.5,-2.5)$. The different curves compare the results for a linear ramp (green lines) with the LM control approach (red lines) and in the absence (solid lines) or presence (dashed lines) of additional CD control fields.}
        \label{fig: Pichler_CD}
    \end{figure}

    In Fig.~\ref{fig: Pichler_CD}(a) and (b) we plot the state preparation fidelity $\mathcal{F}$ and the concurrence of the reduced state of the first two qubits, $\mathcal{C}_{12}$, as a function of the preparation time $T$. In both plots, we compare a linear sweep of $\Omega(t)$ and the LM pulse $\Omega^{\rm LM}(t)$ with and without additional CD control. From those results we find that CD control always improves the fidelity, although for large preparation times the added benefit is minimal. This behavior is expected, as local control fields per se cannot increase entanglement. Still, for short times, $\gamma T\sim 1$, the combination of local qubit rotations and collective dissipation can increase the entanglement compared to what is possible when using only the original driving field, $\Omega(t)$. Specifically, CD driving generates a finite amount of entanglement for arbitrarily short preparation times, which is not the case using LM alone, at the expense of CD amplitudes $\beta_j\propto\dot{\Omega}$ that grow as $T\rightarrow 0$ (see Appendix~\ref{app: CD_Dimer}). This can be understood by looking at the case of $N=2$ qubits, where the action of $H_{\rm CD}$ in Eq.~\eqref{eq: local_CD_sigma_y} on the initial ground state yields (see Appendix~\ref{app: CD_Dimer})
    \begin{align}
        e^{-i\int_0^{T} dt' H_{\rm CD}(t')}\ket{gg} &\simeq  \frac{\ket{gg} - \ket{ee}}{2} - \frac{\ket{S}}{\sqrt{2}}.
    \end{align}
    This state is still factorizable but becomes entangled as soon as a tiny fraction of the state $|ee\rangle$ decays. Therefore, we conclude that single-body CD drive mainly optimizes the initial starting point rather than the long-time decay dynamics.

    \section{Conclusions and outlook}\label{sec: Conclusions}

    In summary, we have shown that the tradeoff between the generated amount of entanglement 
    and preparation time in dissipative state-engineering protocols can be overcome through time-dependent variations of the system's Liouvillian. While even simple linear sweeps of the control parameters can already result in a substantial reduction of the preparation time, we have introduced LM as a general optimization framework for systematically enhancing these and related reservoir-engineering applications. Using this method, we have identified simple analytical pulse shapes that render the preparation  of highly entangled states in distributed waveguide scenarios essentially independent of the target degree of entanglement, yielding speedups of several orders of magnitude. For realistic experimental platforms, where residual decay and dephasing are unavoidable, such dramatic reductions in preparation time are crucial for achieving  the necessary high fidelities required for quantum information processing applications.

    Beyond specific entanglement distribution schemes, we have shown that the LM method provides a general framework for quantum control. It unifies steering and CD control techniques previously developed for isolated quantum systems and extends these ideas to open quantum systems. While this technique is particularly powerful for steering quantum systems along pure steady states, where often analytical results can be obtained, this is not a fundamental limitation. In particular, by evaluating the loss function or the resulting Euler-Lagrange equations numerically, the method can also be applied to accelerate the preparation of mixed steady states or to address more complex multi-parameter control problems. For such applications, the LM method is expected to outperform brute-force optimal-control approaches, especially in large systems where numerical integration of the full dynamics is no longer feasible. This is a situation that frequently arises in reservoir-engineering schemes.


    \begin{acknowledgements}
    We thank Philippe Gigon, Joan Agusti, Rahul Trivedi, and Niklas Glaser for stimulating discussions. This work was supported by the Swiss National Science Foundation through Project No. CRSII 222812/1. A. P.-R. acknowledges support from the European Union’s Marie Skłodowska-Curie Actions under grant agreement No. 101204967 (FTMcQED). This research is part of the Munich Quantum Valley, which is supported by the Bavarian state government with funds from the Hightech Agenda Bayern Plus. M.L. acknowledges support from the Deutsche Forschungsgemeinschaft (DFG, German Research Foundation) under Germany's Excellence Strategy -- EXC-2111 -- 390814868.
    \end{acknowledgements}
    \section{Statement on AI use}
    The authors have utilized Anthropic's Claude Opus for assistance with writing code, the results of which were cross-verified against the human-written code. OpenAI's ChatGPT and Claude Opus were used to verify the numerical results in the preliminary versions of the manuscript. The authors take full responsibility for all results presented in this manuscript.
    
    
    \appendix
    
    \section{Loss function for pure steady states}
    \label{app:Loss_Pure_States}
    In this appendix we outline the derivation of the loss functions in Eq.~\eqref{eq:LossFunction_Pure} and Eq.~\eqref{eq:LossFunction_CD_Pure}, which apply to families of pure steady states and when CD control is coherent. In this case, we can express the loss function as 
    \begin{align}
        \mathscr{L}(\boldsymbol{\lambda}, \boldsymbol{\dot{\lambda}}, \beta) &= \frac{1}{4} \Big\lVert \partial_t \rho_{\rm{ss}}(t) +i[H_{\rm CD}, \rho_{\rm{ss}}(t)]  \Big\rVert_{\rm{tr}}^2\nonumber \\
        & = \frac{1}{4}\left(\mathrm{Tr}\left\{\sqrt{A^\dagger A}\right\}\right)^2,
        \label{eq: Loss_coh}
    \end{align}
    where we introduced the operator $A=|\psi^\prime\rangle\bra{\psi_D}+{\rm H.c.}$ and defined   
    $|\psi^\prime\rangle= \left(\partial_t+i H_{\rm CD}\right) |\psi_D\rangle$, omitting the dependencies on $\boldsymbol{\lambda}$ and $\boldsymbol{\beta}$ for simplicity. We assume that the CD Hamiltonian $H_{\rm CD}$ Hermitian and linear in the vector of control pulses, $\boldsymbol{\beta}$. Since $A^\dag =A$ and its support is restricted to a two dimensional subspace spanned by $|\psi_D\rangle$ and $|\psi^\prime\rangle$, we can write $\sqrt{A^\dag A}=\sqrt{A^2}$ and use the Cayley-Hamilton theorem to obtain
    \begin{equation}
    A^2= {\rm Tr}\{A\} A - \mathbbm{1}_2 \det(A).
    \end{equation}
    The overlap $\bra{\psi_D}\psi^\prime\rangle$ is purely imaginary. Therefore, ${\rm Tr}\{A\}=2{\rm Re} \bra{\psi_D}\psi^\prime\rangle=0$ and 
    \begin{equation}
    \det(A)= |\bra{\psi^\prime}\psi_D\rangle|^2 - \bra{\psi^\prime}\psi^\prime\rangle.
    \label{eq: Det_A}
    \end{equation}
    Substituting Eq.~\eqref{eq: Det_A} into Eq.~\eqref{eq: Loss_coh}, the loss function can be written as
    \begin{align}\label{eq:Loss_Pure_Norm}
          \mathscr{L}(\boldsymbol{\lambda}, \dot{\boldsymbol{\lambda}}, \boldsymbol{\beta}) &= \Big\lvert\big(\mathbbm{1}- \ket{\psi_D}\bra{\psi_D}\big)\big(\partial_t +i H_{\rm CD}\big)\ket{\psi_D} \Big\rvert^2 \nonumber \\
          &=\Big\lvert\big(\mathbbm{1}- \ket{\bar \psi_D}\bra{\bar \psi_D}\big)\partial_t\ket{\bar\psi_D} \Big\rvert^2,
    \end{align}
    where we introduced the modified state 
    \begin{align}
       \ket{\bar\psi_D} = \mathcal{V}(t) \ket{\psi_D}   \equiv  \mathcal{T}_-\exp\left(i\int^t_0 H_{\rm CD}(t') dt' \right)\ket{\psi_D} .
    \end{align} 
    Here, $\mathcal{T}_-$ denotes the antichronological time-ordering, i.e., the unitary operator $\mathcal{V}$  satisfies
    \begin{align}
        \partial_t \mathcal{V} = i \mathcal{V} H_{\rm CD}(t)
    \end{align}
    with $\mathcal{V}(t=0) = \mathbbm{1}$.

    Since the expression in Eq.~\eqref{eq:Loss_Pure_Norm} does not depend on $\dot \beta$, we can solve $\partial \mathscr{L}/\partial \beta = 0$ for $\beta(\blambda, \dot{\blambda})$. The loss function can then be identified with the time-component $\bar{g}_{t,t}$ of the quantum metric tensor~\cite{Provost_80} associated with the family of states $\ket{\bar \psi_D}$, i.e.,
    \begin{align}
    \label{eq:bar_gtt_Appendix}
    \begin{aligned}
     \bar{g}_{t,t} &= \partial_t\bra{\bar{\psi}_D} \big(\mathbbm{1}- \ket{\bar{\psi}_D}\bra{\bar{\psi}_D}\big)\partial_t\ket{\bar{\psi}_D}\\
    &= \sum_{\mu,\nu} \bar{g}_{\mu,\nu} \dot{\lambda}_\mu \dot{\lambda}_\nu,   
    \end{aligned}
    \end{align}
    where we have defined
    \begin{align}
        \bar{g}_{\mu,\nu} = {\rm Re}\!\left\{
    \partial_{\lambda_\mu}\bra{\bar\psi_D}\left(\mathbbm{1}-\ket{\bar\psi_D}\bra{\bar\psi_D}\right)\partial_{\lambda_\nu}\ket{\bar\psi_D}
    \right\}.
    \end{align}
    Note that the second equality in Eq.~\eqref{eq:bar_gtt_Appendix} assumes that the rotated state depends only on $\boldsymbol{\lambda}$ up to a global phase. In the absence of CD controls, $\mathscr{L}$ reduces to the quantum metric tensor $g_{t,t}$ of the original family of states, $\ket{\psi_D}$. Since the process of LM also involves optimizing the CD terms, in general, we have $\bar{g}_{t,t} \leq g_{t,t}$. This method therefore serves as a direct generalization of the geodesic steering scheme of Ref.~\cite{Tomka_16}, where only the \emph{intrinsic} control parameters $\boldsymbol{\lambda}$ are optimized.

    \section{Invariance of the loss function along leakage-minimizing trajectories}\label{app:L_Invariance}
    
    For a loss function of the form $\mathscr L =\bar g_{t,t}$ with $\bar g_{t,t}$ given in Eq.~\eqref{eq:bar_gtt_Appendix}, the corresponding Hamiltonian function
    \begin{align}
        \mathscr H = \sum_\mu \dot{\lambda}_\mu \frac{\partial \mathscr{L}}{\partial \dot{\lambda}_\mu} - \mathscr{L} = \mathscr{L}
    \end{align}
    is equal to the loss function itself. Further, since $\mathscr L =\mathscr H$ does not explicitly depend on time, we can use Noether's theorem \cite{Landau_76} to show that along LM trajectories, the loss function is a constant, i.e., 
    \begin{equation}
    \mathscr{L}\left(\boldsymbol{\lambda}^{\text{LM}}(t), \dot{\boldsymbol{\lambda}}^{\text{LM}}(t)\right)=\rm{const}.
    \end{equation}
    For a single control parameter with
    $\mathscr{L}=\dot{\lambda}^2\bar g_{1,1}(\lambda)$, this invariance implies that the function $\mathcal{I}(\lambda^{\rm LM}(t))= t \times\mathcal{I}(\lambda_{\rm targ})/T$ grows
    linearly in time, where 
    \begin{equation}
    \mathcal{I}(\lambda)
    =\int_{\lambda(0)}^{\lambda} d\lambda^\prime\,
    \sqrt{\bar g_{1,1}(\lambda^\prime)}.
    \end{equation}
    The general solution for the optimized control field is therefore given by
    \begin{equation}
    \lambda^{\rm LM}(t)
    =\mathcal{I}^{-1}\left(
    \mathcal{I}(\lambda_{\rm targ})\frac{t}{T}
    \right).
    \end{equation}
    In the simple cases considered in
    Sec.~\ref{subsec: LM_DD_ent_gen} and
    Sec.~\ref{sec: TMS_ent_gen},
    the integral $\mathcal{I}(\lambda)$ can be expressed in terms
    of an arctangent function. In the general cases, it must be
    evaluated and inverted numerically.

    \section{Time-entanglement tradeoffs}
    \label{app: T-E_bound}
    In all the examples considered in this work involving two qubits only, the master equation can be written as
    \begin{equation}
        \partial_t \rho(t) = -i[H, \rho(t)] +\sum_\mu \mathcal{D}[L_\mu] \rho(t),
    \end{equation}
    with an arbitrary Hamiltonian $H$ and jump operators of the form 
    \begin{equation}
     L_\mu = A_\mu\otimes \mathbbm{1} +  \mathbbm{1} \otimes B_\mu.
    \end{equation}
    For such systems, the bound on the rate of growth of the state preparation fidelity derived in Ref.~\cite{Pocklington_24} takes the form~\footnote{Here, an additional factor of $\sqrt{2}$, which was missing in the derivation of Ref.~\cite{Pocklington_24}, has been accounted for.}
    \begin{equation}\label{eq:Gamma_Bound_App}
        \dot{\mathcal{F}}(t) \leq \Gamma_{\rm{\max}} =\frac{2(d-1)}{p_{\rm{min}}}\sum_\mu|A_\mu|^2 (1-\mathcal{E}_2).
    \end{equation}
    Here, $d=2$ for a two-level system, $\mathcal{E}_2$ is defined in Eq.~\eqref{eq: Scl_ent}, $p_{\rm{min}}$ is the smallest squared Schmidt coefficient in the Schmidt decomposition of the dark state $\ket{\psi_D}$, and $|A_\mu| = \max_{i,j}| (A_\mu)_{i,j}|$ denotes the largest absolute matrix element of the operator $A_\mu$ in the local Schmidt basis.
    
    \subsection{1D waveguide}
    From the reduced density matrix of the two-qubit dark state in Eq.~\eqref{eq: Dark_2qbit}, we obtain
    \begin{equation}
       p_{\rm{min}} = \frac{|\alpha|^2 -\sqrt{2|\alpha|^2 +1} +1}{2(|\alpha|^2 +1)} 
    \end{equation}
    and the scaled entanglement is given by 
    \begin{align}
        \mathcal{E}_2 = 2\left[1 - p^2_{\rm{min}} - (1-p_{\rm{min}})^2\right] = \frac{|\alpha|^4}{(1+|\alpha|^2)^2}.
    \end{align}
    Together with $\lim_{\alpha \to \infty} |A| = \sqrt{(\gamma_L +\gamma_R)}/2$, we obtain the result for the bound in Eq.~\eqref{eq: Pow_law_Q}. 
    
    \subsection{Two-mode squeezed reservoir}
    For the TMS reservoir considered in Sec.~\ref{sec: TMS_ent_gen}, the dark state given in  Eq.~\eqref{eq: Dark_2Qbit_TMS} is already in the Schmidt-decomposed form, from which we obtain $p_{\rm{min}} = \sinh^2(r)/\cosh(2r)$ and $\mathcal{E}_2 = \tanh^2(2r)$. In this setup, however, the jump operators also depend on the squeezing parameter and  
    \begin{align}
        |A_1| = \sqrt{\gamma}\cosh(r), \quad  |A_2| = \sqrt{\gamma}\sinh(r).
    \end{align}
    For large $r$, the bound given in Eq.~\eqref{eq:Gamma_Bound_App} then evaluates to
    \begin{align}
        \Gamma_{\rm{max}} \simeq 4\gamma\sqrt{1-\mathcal{E}_2},
    \end{align}
    with an overall weaker scaling on the entanglement. Within the same parameter regime, in our numerical simulations, we find that the relaxation dynamics at long times is well approximated by  
    \begin{align}
    \begin{split}
            \mathcal{F}(t)\approx &  \bra{\psi_D} \rho(0)\ket{\psi_D}\\
        &+\left(1-\bra{\psi_D} \rho(0)\ket{\psi_D}\right)\left(1-e^{-\gamma_{\rm rel} t}\right),
    \end{split}
     \end{align}
    where $\gamma_{\rm rel}\simeq \lambda_{\rm dis}$ matches the dissipative gap of a TMS reservoir, $\lambda_{\rm dis}= \gamma[6\cosh(2r) -\sqrt{18\cosh(4r) -14}]/4$~\cite{Agusti_23}. For large $r$ and by expressing $\cosh(2r) = 1/\sqrt{1-\mathcal{E}_2}$ and $\cosh(4r) = 2/(1-\mathcal{E}_2) - 1$,
    we find that for $r\gtrsim 2$,
    \begin{align}
         \lambda_{\rm dis}\simeq \frac{2\gamma\sqrt{1-\mathcal{E}_2}}{3}.
         \label{eq: rel_rate_TMS}
     \end{align}
    Therefore, also in this system, $\Gamma_{\rm max}$ predicts the correct entanglement scaling, but is not a tight bound.

    \section{Loss function for two-component states}
    \label{app: Qbit_like}
    In most of the examples considered in this paper, the dark state $\ket{\psi_D(\lambda)}\equiv \ket{\Psi(\lambda)}$ is a two-component state of the form
    \begin{align}
        \ket{\Psi(\lambda)} = \cos(\theta(\lambda)) \ket{\phi_1} + \sin(\theta(\lambda)) e^{i\phi(\lambda)}\ket{\phi_2},
        \label{eq: Qbit_like_st}
    \end{align} 
    where $ \braket{\phi_1 |\phi_2}=0$ and $\partial_\lambda\phi(\lambda) = 0$. In this case 
    the loss function evaluates to 
    \begin{align}
    \begin{split}
       \mathscr{L}(t) =[\partial_\lambda\theta(\lambda)]^2\dot{\lambda}^2,
       \label{eq: loss_2comp}
    \end{split}   
    \end{align}
    and the solution for the corresponding LM control parameter can be expressed in a general form as  
    \begin{align}
    \lambda^{\text{LM}}(t) &= \theta^{-1}\left[\theta(\lambda(0)) + \Big\{\theta(\lambda(T)) - \theta(\lambda(0))\Big\}\frac{t}{T}\right].
    \label{eq: two_comp_LM}
    \end{align}
    This explains why the optimized pulse for the driving field considered in Sec.~\ref{sec: DD_ent_gen} and the squeezing parameter in Sec.~\ref{sec: TMS_ent_gen} have a similar shape. 
    
    In a more general scenario, where the dark state $\ket{\psi_D}$ is a tensor product of many dimers, each of which is of the two-component form of Eq.~\eqref{eq: Qbit_like_st}, we have
    \begin{align}
        &\partial_\lambda \ket{\psi_D(\lambda)} = \sum_{j=1}^{N/2} \ket{\Psi_{1,2}(\lambda)} \otimes \dots \nonumber\\
        &\dots \otimes \partial_\lambda \ket{\Psi_{2j-1,2j}(\lambda)}  \otimes \dots \otimes\ket{\Psi_{N-1,N}(\lambda)}.
        \label{eq: Rt_chng_2comp}
    \end{align}
    The corresponding loss function involves projecting out the component of Eq.~\eqref{eq: Rt_chng_2comp} along $\ket{\psi_D}$, and generally evaluates to
    \begin{align}
        \mathscr{L}(\lambda, \dot{\lambda}) =\dot{\lambda}^2\sum_{j=1}^{N/2} \Big|\Big(\mathbbm{1} - \mathbbm{P}_{2j-1,2j}\Big)\partial_\lambda \ket{\Psi_{2j-1,2j}(\lambda)}\Big|^2, 
    \end{align}
    where $\mathbbm{P}_{2j-1,2j}=\ket{\Psi_{2j-1,2j}(\lambda)}\bra{\Psi_{2j-1,2j}(\lambda)}$. This is just the sum of the loss functions of each two-component dimer.

    \section{Gradient optimization of analytic controls}
    \label{app: GOAT}
    \begin{figure}
        \centering
        \includegraphics[width=\columnwidth]{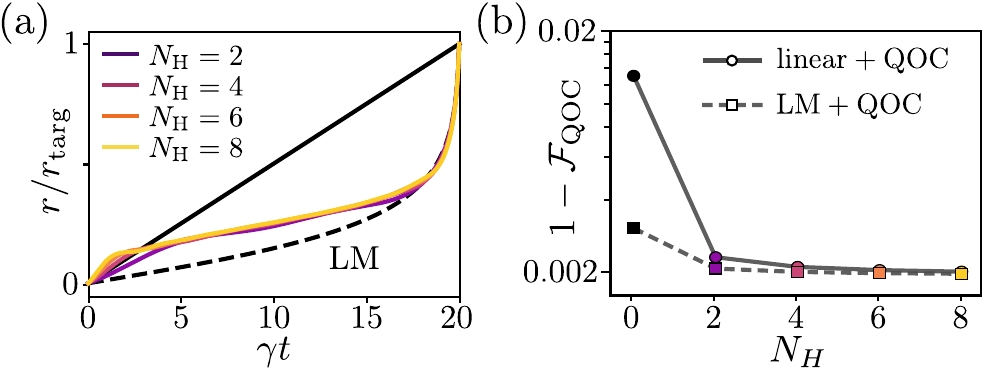}
        \caption{Quantum optimal control for preparing a two-qubit Bell state in a TMS reservoir, as described by Eq.~\eqref{eq: ME_TMS_NQbit} for $N_p=1$ and $\gamma_\phi=0$. (a) Plot of the optimized pulse shape $r(t)$ obtained from QOC for different numbers of harmonics $N_H$ in the expansion in Eq.~\eqref{eq: QOC_ansatz} and using a linear ramp (black solid line) as a seed. The dashed line indicates the shape of the LM pulse, $r^{\rm LM}(t)$. (b) The final state preparation fidelity $\mathcal{F}_{\rm QOC}$ is plotted as a function of the number of harmonics and using either the linear sweep (solid line) or the LM pulse (dashed line) as a seed. For both plots, we have assumed a final squeezing strength of $r_{\rm targ} = 3$ and $\delta_{1,A}=\delta_{1,B}=0$. }
        \label{fig: QOC_TMS}
    \end{figure} 
    Following the arguments presented in Ref.~\cite{Machnes_18} for isolated quantum systems, we find that for an open quantum system with Liouvillian $\mathcal{L}[\boldsymbol{x}]$, where $\boldsymbol{x}=(x_1,x_2,\dots)$ is a set of variational parameters, the gradient of the density operator, $\partial \rho[\boldsymbol{x}]/\partial x_n$, evolves as  
    \begin{align}
    \begin{split}
        \partial_t \left(\frac{\partial  \rho[\boldsymbol{x}]}{ \partial x_n} \right)&= \frac{\partial }{\partial x_n}  \left(\mathcal{L}[\boldsymbol{x}] \rho[\boldsymbol{x}]\right) \\
        &=\frac{\partial \mathcal{L}[\boldsymbol{x}]}{\partial x_n}\rho[\boldsymbol{x}] + \mathcal{L} \frac{\partial \rho[\boldsymbol{x}]}{\partial x_n}.
        \label{eq: QOC_Machnes}
    \end{split}
    \end{align}
    Therefore, in order to evaluate the gradient of the cost function $\zeta =1- \bra{\psi_D(T)} \rho(T) \ket{\psi_D(T)}$, the components $\partial \rho(T)/\partial x_n$ can be evaluated by integrating Eq.~\eqref{eq: QOC_Machnes} together with the master equation for $\rho$. This increases the total dimension of the problem, approximately proportional to the number of optimizable parameters, but provides the optimizer with numerically very accurate estimates for the gradient. 
    
    In Fig.~\ref{fig: QOC_TMS}, we perform an analysis similar to that in Sec.~\ref{subsec: QOC_DD_ent_gen}, but for the preparation of a two-qubit Bell state in a TMS environment and using the ansatz
    \begin{align}
    \begin{split}
        r(t) =\, r_{\rm seed}(t) 
        + \sum_{n=1}^{N_H}\Big( &a_n \sin\left(\omega_n t\right) \\
        &+ b_n \left[\cos\left(\omega_n  t\right)-1\right]\Big)
    \end{split}
    \end{align}
    for the squeezing parameter. The general trends and the performance of the optimizer for different seed functions are similar as for the example shown in Fig.~\ref{fig: QOC_Stannigel}.

    \section{Counterdiabatic driving for the dimerized steady state}
    \label{app: CD_Dimer}
    In this appendix, we provide additional details on the CD driving scheme investigated in 
    Sec.~\ref{subsec:CD driving}. In this example, we consider a bidirectional waveguide with $\gamma_A=0$ and detuning pattern satisfying $\delta_{2k-1}=- \delta_{2k}$, which results in a dimerized steady state of the form  
    \begin{align}
        |\psi_D\rangle  =  \bigotimes_{j=1}^{N/2} \left( \frac{\ket{gg} -\alpha_j \ket{S}}{\sqrt{1+ \alpha^2_j}}\right),
        \label{eq: Bidirec_Dim_Dark}
    \end{align}
    where $\alpha_j=\sqrt{2}\Omega/\delta_{2j-1}$. 
    
    For $N$ qubits, the most general local CD control Hamiltonian can be expressed as 
    \begin{align}
       H_{\rm CD}[\boldsymbol{\beta}(t)] = \sum_{j=1}^N\left[ \beta^x_j(t) \sigma^x_j + \beta^y_j (t)\sigma^y_j + \beta^z_j(t)\sigma^z_j \right].
    \end{align} 
    However, when taking a look at the term $(\partial_t+i H_{\rm CD})|\psi_D\rangle$ in the expression of the loss function in Eq.~\eqref{eq:LossFunction_CD_Pure}, we see that only terms proportional to $\sigma^y=i(|e\rangle\langle g|- |g\rangle\langle e|)$ can contribute to a partial cancellation of nonadiabatic transitions. Therefore, we can set $\beta^x_j(t)=\beta_j^z(t)=0$ and identify $\beta_j(t)\equiv\beta^y_j(t)$ from now on. 
    
    \subsection{Single dimer}
    For the case of a single dimer with $N=2$ and $\delta_1=-\delta_2=\delta$, the loss function reads
    \begin{align}
    \mathscr{L} = \Big|\Big|\partial_t \ket{D(\alpha)}  + i\left(H_{\rm{CD}}-\braket{ H_{\rm{CD}} }\right)\ket{D(\alpha)}\Big|\Big|^2,
    \label{eq: loss_CD_Pichler_N2}
    \end{align}
    where $\ket{D(\alpha)}=(\ket{gg} -\alpha\ket{S})/\sqrt{1+\alpha^2}$ and $\braket{H_{\rm{CD}}} = \braket{D(\alpha)| H_{\rm{CD}} |D(\alpha)}$. We see that CD driving adds an additional term in the loss function, where $\braket{H_{\rm{CD}}}=0$ and 
    \begin{align}
    \begin{split}
       iH_{\rm{CD}}\ket{D(\alpha)}  
       = & -\frac{ (\beta_1\ket{eg} +\beta_2\ket{ge})}{\sqrt{1+ \alpha^2}}\\&- \frac{\alpha(\beta_1-\beta_2)(\ket{gg} +\ket{ee})}{\sqrt{2(1 + \alpha^2)}}.
        \label{eq: T1_CD}
    \end{split}
    \end{align}
    By defining $\beta_\pm=(\beta_1\pm \beta_2)/2$, we see that the contribution $\sim \beta_+$ in this expression is orthogonal to $\partial_t \ket{D(\alpha)}$ and cannot reduce the cost function. Therefore, setting $\beta_+=0$ and identifying $\beta_-(t)\equiv \beta(t)$, the loss function is given by
    \begin{equation}
        \mathscr{L}(\Omega, \dot{\Omega}, \beta) = \left(\frac{\dot{\alpha}}{1+\alpha^2} + \sqrt{2}\beta\right)^2 + \frac{2\beta^2 \alpha^2}{1+\alpha^2}.
        \label{eq: Pichler_CD_loss}
    \end{equation}
    This function is independent of $\dot{\beta}$, and from the resulting Euler-Lagrange equation, $\partial \mathscr L/\partial \beta =0 $, we can express the optimal CD control field as
    \begin{align}
        \beta(t) = \frac{-\dot{\Omega}(t)\delta}{\delta^2 + 4\Omega^2(t)}.
        \label{eq: beta_CD}
    \end{align}
    By substituting this result back into Eq.~\eqref{eq: Pichler_CD_loss}, we obtain the effective loss function
    \begin{equation}
        \begin{split}
            \mathscr{L}_{\rm eff}(\alpha,\dot\alpha) =\frac{\dot{\alpha}^2 \alpha^2}{(1+\alpha^2)^2(1+2\alpha^2)}.
        \end{split}
        \label{eq: loss_CD_w_beta}
    \end{equation}
    From this expression, the optimal control parameter $\Omega^{\rm LM}(t)=\alpha^{\rm LM}(t)\delta/\sqrt{2}$ can be obtained from the general solution derived in Appendix~\ref{app:L_Invariance} and inserted back into Eq.~\eqref{eq: beta_CD} to obtain the corresponding optimal CD driving strength.   
    
    Since $\beta \sim\dot{\Omega}$, the CD driving introduces a non-trivial rotation of the two-qubit state, even at very small protocol times. Specifically,  for $\Omega(0)=0$ and $\Omega(T)\gg |\delta|$, the unitary rotation generated by $H_{\rm CD}(t)$,  
    \begin{align}
    \mathcal{U} = e^{-i\int_0^{T} dt' \beta(t') (\sigma^y_1 - \sigma^y_2)}\simeq U_{-\frac{\pi}{2}}\otimes U_{\frac{\pi}{2}},
    \end{align}
    corresponds to a $\pi/2$-rotation of each qubit. Therefore, when applied to the initial state,
    \begin{align}
        \mathcal{U} \ket{gg} &= \left(\frac{\ket{g}-\ket{e}}{\sqrt{2}}\right)\left(\frac{\ket{g}+\ket{e}}{\sqrt{2}}\right)   \\
        &= \frac{\ket{gg} - \ket{ee}}{2} - \frac{\ket{S}}{\sqrt{2}},
    \end{align}
    the CD control rotates the qubits into a state that is still a product, but has the largest overlap with the singlet state. This tendency is also observed for the full dynamics and explains why local driving fields can still assist in the generation of entanglement. Note that the pulse area  $\int_0^t dt' \beta(t')=-\frac{1}{2}\arctan[2\Omega(t)/\delta]$ stays finite although $\beta\propto \dot{\Omega}$ diverges as $t^{-\frac{1}{2}}$ for $t\rightarrow 0$ (the CD-optimized pulse initially grows as $\Omega^{\rm LM}\propto \sqrt{t}$, since $\mathscr{L}_{\rm eff}\propto \alpha^2$). Numerically, $\dot{\Omega}$ in Eq.~\eqref{eq: beta_CD} is evaluated as a finite difference over the time bin, which reproduces this area as the time bin is reduced.

    \subsection{Multiple dimers}
    Since the steady state in Eq.~\eqref{eq: Bidirec_Dim_Dark} can be written as a product of dimers, the analysis from above can be readily generalized to arbitrary (even) numbers of qubits by evaluating the action of $\partial_t+iH_{\rm CD}$ on each dimer separately. By restricting the CD driving to the form   
    \begin{align}
        H_{\rm{CD}}(t) = \sum_{j=1}^{N} \beta_j(t)\sigma^y_{j}
    \end{align}
    and following the same steps as above, we obtain the effective loss function
    \begin{align}\label{eq: Loss_eff_CD}
        \mathscr{L}_{\rm eff}(\Omega,\dot\Omega) = \sum_{j=1}^{N/2} \frac{\alpha^2_{j} \dot{\alpha}_{j}^2}{(1+\alpha_{j}^2)^2 (1+2\alpha_{j}^2)}.
    \end{align}
    Here, we have already implemented the constraint $\partial \mathscr{L}/\partial \beta_j = 0$ to eliminate the CD controls using Eq.~\eqref{eq: beta_j_CD} in the main text. The optimal parameters $\Omega^{\rm LM}(t)$ and $\beta_j(t)$ can again be derived from the general solution in Appendix~\ref{app:L_Invariance}.

    \bibliographystyle{apsrev4-2}
    \bibliography{bibliography}

@article{Goll_26,
  doi = {10.22331/q-2026-07-15-2160},
  url = {https://doi.org/10.22331/q-2026-07-15-2160},
  title = {Geometric speed limit of state preparation and curved control spaces},
  author = {Goll, Maximilian and Jonsson, Robert H.},
  journal = {{Quantum}},
  issn = {2521-327X},
  publisher = {{Verein zur F{\"{o}}rderung des Open Access Publizierens in den Quantenwissenschaften}},
  volume = {10},
  pages = {2160},
  month = jul,
  year = {2026}
}

@misc{Dengis_26,
      title={Time-independent counterdiabatic driving for emergent two-level subspaces in many-body systems}, 
      author={Simon Dengis and Peter Schlagheck},
      year={2026},
      eprint={2607.12848},
      archivePrefix={arXiv},
      primaryClass={quant-ph}
}

@book{Milnor_63,
  author    = {Milnor, John},
  title     = {{Morse Theory}},
  series    = {Annals of Mathematics Studies},
  volume    = {51},
  publisher = {Princeton University Press},
  address   = {Princeton, NJ},
  year      = {1963},
  isbn      = {9780691080086},
  doi       = {10.1515/9781400881802},
  note      = {Based on lecture notes by M. Spivak and R. Wells}
}

@book{Docarmo_92,
  author    = {do Carmo, Manfredo P.},
  title     = {{Riemannian Geometry}},
  series    = {Mathematics: Theory \& Applications},
  publisher = {Birkh\"{a}user},
  address   = {Boston, MA},
url = {https://link.springer.com/book/9780817634902},
  year      = {1992},
  isbn      = {9780817634902},
}

@article{Grabarits_26,
  title = {Fighting Exponentially Small Gaps by Counterdiabatic Driving},
  author = {Grabarits, Andr\'as and Balducci, Federico and del Campo, Adolfo},
  journal = {PRX Quantum},
  volume = {7},
  issue = {1},
  pages = {010322},
  numpages = {43},
  year = {2026},
  month = {Feb},
  publisher = {American Physical Society},
  doi = {10.1103/tgzt-dy3h},
  url = {https://link.aps.org/doi/10.1103/tgzt-dy3h}
}

@article{Cepaite_23,
  title = {Counterdiabatic Optimized Local Driving},
  author = {\ifmmode \check{C}\else \v{C}\fi{}epait\ifmmode \dot{e}\else \.{e}\fi{}, Ieva and Polkovnikov, Anatoli and Daley, Andrew J. and Duncan, Callum W.},
  journal = {PRX Quantum},
  volume = {4},
  issue = {1},
  pages = {010312},
  numpages = {21},
  year = {2023},
  month = {Jan},
  publisher = {American Physical Society},
  doi = {10.1103/PRXQuantum.4.010312},
  url = {https://link.aps.org/doi/10.1103/PRXQuantum.4.010312}
}

@article{Stannigel_12,
doi = {10.1088/1367-2630/14/6/063014},
url = {https://doi.org/10.1088/1367-2630/14/6/063014},
year = {2012},
month = {jun},
publisher = {IOP Publishing},
volume = {14},
number = {6},
pages = {063014},
author = {Stannigel, K and Rabl, P and Zoller, P},
title = {Driven-dissipative preparation of entangled states in cascaded quantum-optical networks},
journal = {New J. Phys.},
}

@article{Agusti_22,
  title = {Long-distance distribution of qubit-qubit entanglement using Gaussian-correlated photonic beams},
  author = {Agust\'{\i}, J. and Minoguchi, Y. and Fink, J. M. and Rabl, P.},
  journal = {Phys. Rev. A},
  volume = {105},
  issue = {6},
  pages = {062454},
  numpages = {14},
  year = {2022},
  month = {Jun},
  publisher = {American Physical Society},
  doi = {10.1103/PhysRevA.105.062454},
  url = {https://link.aps.org/doi/10.1103/PhysRevA.105.062454}
}

@article{Avron_12,
   title={Adiabatic Theorems for Generators of Contracting Evolutions},
   volume={314},
   ISSN={1432-0916},
   url={http://dx.doi.org/10.1007/s00220-012-1504-1},
   DOI={10.1007/s00220-012-1504-1},
   number={1},
   journal={Commun. Math. Phys.},
   publisher={Springer Science and Business Media LLC},
   author={Avron, J. E. and Fraas, M. and Graf, G. M. and Grech, P.},
   year={2012},
   month=may, 
   pages={163},
}

@article{Vacanti_2014,
doi = {10.1088/1367-2630/16/5/053017},
url = {https://doi.org/10.1088/1367-2630/16/5/053017},
year = {2014},
month = {may},
publisher = {IOP Publishing},
volume = {16},
number = {5},
pages = {053017},
author = {Vacanti, G and Fazio, R and Montangero, S and Palma, G M and Paternostro, M and Vedral, V},
title = {Transitionless quantum driving in open quantum systems},
journal = {New J. Phys.}
}

@article{Campo_13,
  title = {Shortcuts to Adiabaticity by Counterdiabatic Driving},
  author = {del Campo, Adolfo},
  journal = {Phys. Rev. Lett.},
  volume = {111},
  issue = {10},
  pages = {100502},
  numpages = {5},
  year = {2013},
  month = {Sep},
  publisher = {American Physical Society},
  doi = {10.1103/PhysRevLett.111.100502},
  url = {https://link.aps.org/doi/10.1103/PhysRevLett.111.100502}
}

@article{Berry_09,
doi = {10.1088/1751-8113/42/36/365303},
url = {https://doi.org/10.1088/1751-8113/42/36/365303},
year = {2009},
month = {aug},
publisher = {},
volume = {42},
number = {36},
pages = {365303},
author = {Berry, M V},
title = {Transitionless quantum driving},
journal = {J. Phys. A: Math. Theor.},
}

@article{Duncan_24,
  title = {Exact counterdiabatic driving in finite topological lattice models},
  author = {Duncan, Callum W.},
  journal = {Phys. Rev. B},
  volume = {109},
  issue = {24},
  pages = {245421},
  numpages = {11},
  year = {2024},
  month = {Jun},
  publisher = {American Physical Society},
  doi = {10.1103/PhysRevB.109.245421},
  url = {https://link.aps.org/doi/10.1103/PhysRevB.109.245421}
}

@article{Santos_21,
  title = {Quantum adiabatic brachistochrone for open systems},
  author = {Santos, Alan C. and Villas-Boas, C. J. and Bachelard, R.},
  journal = {Phys. Rev. A},
  volume = {103},
  issue = {1},
  pages = {012206},
  numpages = {9},
  year = {2021},
  month = {Jan},
  publisher = {American Physical Society},
  doi = {10.1103/PhysRevA.103.012206},
  url = {https://link.aps.org/doi/10.1103/PhysRevA.103.012206}
}

@article{Alipour_2020,
   title={Shortcuts to Adiabaticity in Driven Open Quantum Systems: Balanced Gain and Loss and Non-Markovian Evolution},
   volume={4},
   ISSN={2521-327X},
   url={http://dx.doi.org/10.22331/q-2020-09-28-336},
   DOI={10.22331/q-2020-09-28-336},
   journal={Quantum},
   publisher={Verein zur Forderung des Open Access Publizierens in den Quantenwissenschaften},
   author={Alipour, Sahar and Chenu, Aurelia and Rezakhani, Ali T. and del Campo, Adolfo},
   year={2020},
   month=sep, pages={336} }

@article{Shah_24,
  title = {Stabilizing Remote Entanglement via Waveguide Dissipation},
  author = {Shah, Parth S. and Yang, Frank and Joshi, Chaitali and Mirhosseini, Mohammad},
  journal = {PRX Quantum},
  volume = {5},
  issue = {3},
  pages = {030346},
  numpages = {23},
  year = {2024},
  month = {Sep},
  publisher = {American Physical Society},
  doi = {10.1103/PRXQuantum.5.030346},
  url = {https://link.aps.org/doi/10.1103/PRXQuantum.5.030346}
}

@article{Santos_21_TQD,
  title = {Generalized transitionless quantum driving for open quantum systems},
  author = {Santos, Alan C. and Sarandy, Marcelo S.},
  journal = {Phys. Rev. A},
  volume = {104},
  issue = {6},
  pages = {062421},
  numpages = {19},
  year = {2021},
  month = {Dec},
  publisher = {American Physical Society},
  doi = {10.1103/PhysRevA.104.062421},
  url = {https://link.aps.org/doi/10.1103/PhysRevA.104.062421}
}

@article{Lim_24,
  title = {Exponentially faster preparation of quantum dimers via driven-dissipative stabilization},
  author = {Lim, Kian Hwee and Mok, Wai-Keong and You, Jia-Bin and Kong, Jian Feng and Aghamalyan, Davit},
  journal = {Phys. Rev. Res.},
  volume = {6},
  issue = {3},
  pages = {L032047},
  numpages = {6},
  year = {2024},
  month = {Aug},
  publisher = {American Physical Society},
  doi = {10.1103/PhysRevResearch.6.L032047},
  url = {https://link.aps.org/doi/10.1103/PhysRevResearch.6.L032047}
}

@article{Bukov_19,
  title = {Geometric Speed Limit of Accessible Many-Body State Preparation},
  author = {Bukov, Marin and Sels, Dries and Polkovnikov, Anatoli},
  journal = {Phys. Rev. X},
  volume = {9},
  issue = {1},
  pages = {011034},
  numpages = {21},
  year = {2019},
  month = {Feb},
  publisher = {American Physical Society},
  doi = {10.1103/PhysRevX.9.011034},
  url = {https://link.aps.org/doi/10.1103/PhysRevX.9.011034}
}

@article{Hegade_21_PRA,
  title = {Shortcuts to Adiabaticity in Digitized Adiabatic Quantum Computing},
  author = {Hegade, Narendra N. and Paul, Koushik and Ding, Yongcheng and Sanz, Mikel and Albarr\'an-Arriagada, F. and Solano, Enrique and Chen, Xi},
  journal = {Phys. Rev. Appl.},
  volume = {15},
  issue = {2},
  pages = {024038},
  numpages = {13},
  year = {2021},
  month = {Feb},
  publisher = {American Physical Society},
  doi = {10.1103/PhysRevApplied.15.024038},
  url = {https://link.aps.org/doi/10.1103/PhysRevApplied.15.024038}
}

@article{Dann_19,
  title = {Shortcut to Equilibration of an Open Quantum System},
  author = {Dann, Roie and Tobalina, Ander and Kosloff, Ronnie},
  journal = {Phys. Rev. Lett.},
  volume = {122},
  issue = {25},
  pages = {250402},
  numpages = {6},
  year = {2019},
  month = {Jun},
  publisher = {American Physical Society},
  doi = {10.1103/PhysRevLett.122.250402},
  url = {https://link.aps.org/doi/10.1103/PhysRevLett.122.250402}
}

@article{Dupays_20,
  title = {Superadiabatic thermalization of a quantum oscillator by engineered dephasing},
  author = {Dupays, L. and Egusquiza, I. L. and del Campo, A. and Chenu, A.},
  journal = {Phys. Rev. Res.},
  volume = {2},
  issue = {3},
  pages = {033178},
  numpages = {9},
  year = {2020},
  month = {Aug},
  publisher = {American Physical Society},
  doi = {10.1103/PhysRevResearch.2.033178},
  url = {https://link.aps.org/doi/10.1103/PhysRevResearch.2.033178}
}

@article{Hegade_22,
  title = {Portfolio optimization with digitized counterdiabatic quantum algorithms},
  author = {Hegade, N. N. and Chandarana, P. and Paul, K. and Chen, Xi and Albarr\'an-Arriagada, F. and Solano, E.},
  journal = {Phys. Rev. Res.},
  volume = {4},
  issue = {4},
  pages = {043204},
  numpages = {9},
  year = {2022},
  month = {Dec},
  publisher = {American Physical Society},
  doi = {10.1103/PhysRevResearch.4.043204},
  url = {https://link.aps.org/doi/10.1103/PhysRevResearch.4.043204}
}

@article{Hegade_21,
  title = {Digitized adiabatic quantum factorization},
  author = {Hegade, Narendra N. and Paul, Koushik and Albarr\'an-Arriagada, F. and Chen, Xi and Solano, Enrique},
  journal = {Phys. Rev. A},
  volume = {104},
  issue = {5},
  pages = {L050403},
  numpages = {6},
  year = {2021},
  month = {Nov},
  publisher = {American Physical Society},
  doi = {10.1103/PhysRevA.104.L050403},
  url = {https://link.aps.org/doi/10.1103/PhysRevA.104.L050403}
}

@article{Duncan_25,
  title = {Taming Quantum Systems: A Tutorial for Using Shortcuts-To-Adiabaticity, Quantum Optimal Control, and Reinforcement Learning},
  author = {Duncan, Callum W. and Poggi, Pablo M. and Bukov, Marin and Zinner, Nikolaj Thomas and Campbell, Steve},
  journal = {PRX Quantum},
  volume = {6},
  issue = {4},
  pages = {040201},
  numpages = {69},
  year = {2025},
  month = {Oct},
  publisher = {American Physical Society},
  doi = {10.1103/j8c7-v2hd},
  url = {https://link.aps.org/doi/10.1103/j8c7-v2hd}
}

@article{Claeys_19,
  title = {Floquet-Engineering Counterdiabatic Protocols in Quantum Many-Body Systems},
  author = {Claeys, Pieter W. and Pandey, Mohit and Sels, Dries and Polkovnikov, Anatoli},
  journal = {Phys. Rev. Lett.},
  volume = {123},
  issue = {9},
  pages = {090602},
  numpages = {7},
  year = {2019},
  month = {Aug},
  publisher = {American Physical Society},
  doi = {10.1103/PhysRevLett.123.090602},
  url = {https://link.aps.org/doi/10.1103/PhysRevLett.123.090602}
}

@article{Irfan_25,
  title = {Autonomous Stabilization of Remote Entanglement in a Cascaded Quantum Network},
  author = {Irfan, Abdullah and Singirikonda, Kaushik and Yao, Mingxing and Lingenfelter, Andrew and Mollenhauer, Michael and Cao, Xi and Clerk, Aashish A. and Pfaff, Wolfgang},
  journal = {Phys. Rev. X},
  volume = {16},
  issue = {3},
  pages = {031004},
  numpages = {20},
  year = {2026},
  month = {Jul},
  publisher = {American Physical Society},
  doi = {10.1103/z6zz-vw5q},
  url = {https://link.aps.org/doi/10.1103/z6zz-vw5q}
}

@article{Agusti_23,
  title = {Autonomous Distribution of Programmable Multiqubit Entanglement in a Dual-Rail Quantum Network},
  author = {Agust\'{\i}, J. and Zhang, X. H. H. and Minoguchi, Y. and Rabl, P.},
  journal = {Phys. Rev. Lett.},
  volume = {131},
  issue = {25},
  pages = {250801},
  numpages = {7},
  year = {2023},
  month = {Dec},
  publisher = {American Physical Society},
  doi = {10.1103/PhysRevLett.131.250801},
  url = {https://link.aps.org/doi/10.1103/PhysRevLett.131.250801}
}

@article{Pocklington_24,
  title = {Universal Time-Entanglement Trade-Off in Open Quantum Systems},
  author = {Pocklington, Andrew and Clerk, Aashish A.},
  journal = {PRX Quantum},
  volume = {5},
  issue = {4},
  pages = {040305},
  numpages = {27},
  year = {2024},
  month = {Oct},
  publisher = {American Physical Society},
  doi = {10.1103/PRXQuantum.5.040305},
  url = {https://link.aps.org/doi/10.1103/PRXQuantum.5.040305}
}

@article{Ljubotina_22,
  title = {Optimal Steering of Matrix Product States and Quantum Many-Body Scars},
  author = {Ljubotina, Marko and Roos, Barbara and Abanin, Dmitry A. and Serbyn, Maksym},
  journal = {PRX Quantum},
  volume = {3},
  issue = {3},
  pages = {030343},
  numpages = {24},
  year = {2022},
  month = {Sep},
  publisher = {American Physical Society},
  doi = {10.1103/PRXQuantum.3.030343},
  url = {https://link.aps.org/doi/10.1103/PRXQuantum.3.030343}
}

@misc{Tomka_16,
      title={Geodesic Paths for Quantum Many-Body Systems}, 
      author={Michael Tomka and Tiago Souza and Steven Rosenberg and Anatoli Polkovnikov},
      year={2016},
      eprint={1606.05890},
      archivePrefix={arXiv},
      primaryClass={cond-mat.quant-gas}
}

@article{Kraus_08,
  title = {Preparation of entangled states by quantum Markov processes},
  author = {Kraus, B. and B\"uchler, H. P. and Diehl, S. and Kantian, A. and Micheli, A. and Zoller, P.},
  journal = {Phys. Rev. A},
  volume = {78},
  issue = {4},
  pages = {042307},
  numpages = {9},
  year = {2008},
  month = {Oct},
  publisher = {American Physical Society},
  doi = {10.1103/PhysRevA.78.042307},
  url = {https://link.aps.org/doi/10.1103/PhysRevA.78.042307}
}

@article{Kolodrubetz_17,
title = {Geometry and non-adiabatic response in quantum and classical systems},
journal = {Phys. Rep.},
volume = {697},
pages = {1-87},
year = {2017},
issn = {0370-1573},
doi = {https://doi.org/10.1016/j.physrep.2017.07.001},
url = {https://www.sciencedirect.com/science/article/pii/S0370157317301989},
author = {Michael Kolodrubetz and Dries Sels and Pankaj Mehta and Anatoli Polkovnikov},
}

@article{Venuti_16,
  title = {Adiabaticity in open quantum systems},
  author = {Venuti, Lorenzo Campos and Albash, Tameem and Lidar, Daniel A. and Zanardi, Paolo},
  journal = {Phys. Rev. A},
  volume = {93},
  issue = {3},
  pages = {032118},
  numpages = {11},
  year = {2016},
  month = {Mar},
  publisher = {American Physical Society},
  doi = {10.1103/PhysRevA.93.032118},
  url = {https://link.aps.org/doi/10.1103/PhysRevA.93.032118}
}

@article{Verstraete_09,
  author  = {Verstraete, Frank and Wolf, Michael M. and Cirac, J. Ignacio},
  title   = {{Quantum Computation and Quantum-State Engineering Driven by Dissipation}},
  journal = {Nat. Phys.},
  volume  = {5},
  number  = {9},
  pages   = {633--636},
  year    = {2009},
  doi     = {10.1038/nphys1342}
}

@article{Agusti_25,
   title={Non-Markovian thermal reservoirs for autonomous entanglement distribution},
   volume={10},
   ISSN={2521-327X},
   url={http://dx.doi.org/10.22331/q-2026-04-15-2066},
   DOI={10.22331/q-2026-04-15-2066},
   journal={Quantum},
   publisher={Verein zur Forderung des Open Access Publizierens in den Quantenwissenschaften},
   author={Agustí, Joan and Schneider, Christian M. F. and Fedorov, Kirill G. and Filipp, Stefan and Rabl, Peter},
   year={2026},
   month=Apr, pages={2066} }

@book{Nielsen_00,
  author = {Nielsen, Michael A. and Chuang, Isaac L.},
address = {Cambridge},
  publisher = {Cambridge University Press},
  title = {Quantum Computation and Quantum Information},
doi = {10.1017/CBO9780511976667},
  year = 2000
}

@article{Venuti_17,
  title = {Relaxation versus adiabatic quantum steady-state preparation},
  author = {Venuti, Lorenzo Campos and Albash, Tameem and Marvian, Milad and Lidar, Daniel and Zanardi, Paolo},
  journal = {Phys. Rev. A},
  volume = {95},
  issue = {4},
  pages = {042302},
  numpages = {14},
  year = {2017},
  month = {Apr},
  publisher = {American Physical Society},
  doi = {10.1103/PhysRevA.95.042302},
  url = {https://link.aps.org/doi/10.1103/PhysRevA.95.042302}
}

@article{Muller_12,
title = {Engineered Open Systems and Quantum Simulations with Atoms and Ions},
journal = {Adv. At. Mol. Opt. Phy.},
volume = {61},
pages = {1-80},
year = {2012},
doi = {https://doi.org/10.1016/B978-0-12-396482-3.00001-6},
url = {https://www.sciencedirect.com/science/article/pii/B9780123964823000016},
author = {Markus Müller and Sebastian Diehl and Guido Pupillo and Peter Zoller},
}

@article{Caneva_11,
  title = {Chopped random-basis quantum optimization},
  author = {Caneva, Tommaso and Calarco, Tommaso and Montangero, Simone},
  journal = {Phys. Rev. A},
  volume = {84},
  issue = {2},
  pages = {022326},
  numpages = {9},
  year = {2011},
  month = {Aug},
  publisher = {American Physical Society},
  doi = {10.1103/PhysRevA.84.022326},
  url = {https://link.aps.org/doi/10.1103/PhysRevA.84.022326}
}

@article{Petruhanov_2023,
   title={GRAPE optimization for open quantum systems with time-dependent decoherence rates driven by coherent and incoherent controls},
   volume={56},
   ISSN={1751-8121},
   url={http://dx.doi.org/10.1088/1751-8121/ace13f},
   DOI={10.1088/1751-8121/ace13f},
   number={30},
   journal={J. Phys. A: Math. Theor.},
   publisher={IOP Publishing},
   author={Petruhanov, V N and Pechen, A N},
   year={2023},
   month={July}, pages={305303} }

@article{Gautier_25,
  title = {Optimal Control in Large Open Quantum Systems: The Case of Transmon Readout and Reset},
  author = {Gautier, Ronan and Genois, \'Elie and Blais, Alexandre},
  journal = {Phys. Rev. Lett.},
  volume = {134},
  issue = {7},
  pages = {070802},
  numpages = {7},
  year = {2025},
  month = {Feb},
  publisher = {American Physical Society},
  doi = {10.1103/PhysRevLett.134.070802},
  url = {https://link.aps.org/doi/10.1103/PhysRevLett.134.070802}
}

@article{Govia_22,
  author  = {Govia, Luke C. G. and Lingenfelter, Andrew and Clerk, Aashish A.},
  title   = {{Stabilizing Two-Qubit Entanglement by Mimicking a Squeezed Environment}},
  journal = {Phys. Rev. Res.},
  volume  = {4},
  number  = {2},
  pages   = {023010},
  year    = {2022},
  doi     = {10.1103/PhysRevResearch.4.023010}
}

@article{Krauter_11,
  author  = {Krauter, Hanna and Muschik, Christine A. and Jensen, Kasper and Wasilewski, Wojciech
             and Petersen, Jonas M. and Cirac, J. Ignacio and Polzik, Eugene S.},
  title   = {{Entanglement Generated by Dissipation and Steady State Entanglement of Two Macroscopic Objects}},
  journal = {Phys. Rev. Lett.},
  volume  = {107},
  number  = {8},
  pages   = {080503},
  year    = {2011},
  doi     = {10.1103/PhysRevLett.107.080503}
}

@article{Juanes_26,
  title = {Distributing Stationary Qubit Entanglement through a Nonlocal Squeezed Reservoir},
  author = {Andr\'es-Juanes, A. and Agust\'{\i}, J. and Sett, R. and Redchenko, E. S. and Kapoor, L. N. and Hawaldar, S. and Rabl, P. and Fink, J. M.},
  journal = {Phys. Rev. X},
  volume = {16},
  issue = {3},
  pages = {031005},
  numpages = {20},
  year = {2026},
  month = {Jul},
  publisher = {American Physical Society},
  doi = {10.1103/r4jt-j39w},
  url = {https://link.aps.org/doi/10.1103/r4jt-j39w}
}

@article{Pocklington_25,
  title = {Accelerating Dissipative State Preparation with Adaptive Open Quantum Dynamics},
  author = {Pocklington, Andrew and Clerk, Aashish A.},
  journal = {Phys. Rev. Lett.},
  volume = {134},
  issue = {5},
  pages = {050603},
  numpages = {7},
  year = {2025},
  month = {Feb},
  publisher = {American Physical Society},
  doi = {10.1103/PhysRevLett.134.050603},
  url = {https://link.aps.org/doi/10.1103/PhysRevLett.134.050603}
}

@ARTICLE{Provost_80,
       author = {{Provost}, J.~P. and {Vallee}, G.},
        title = "{Riemannian structure on manifolds of quantum states}",
      journal = {Commun. Math. Phys.},
         year = 1980,
        month = sep,
       volume = {76},
       number = {3},
        pages = {289-301},
          doi = {10.1007/BF02193559},
       adsurl = {https://ui.adsabs.harvard.edu/abs/1980CMaPh..76..289P}
}

@misc{Yao_26,
      title={Floquet Reservoir Engineering for Remote Logical Entanglement}, 
      author={Mingxing Yao and Aashish A. Clerk},
      year={2026},
      eprint={2607.21360},
      archivePrefix={arXiv},
      primaryClass={quant-ph}
}

@article{Poyatos_96,
  title = {Quantum Reservoir Engineering with Laser Cooled Trapped Ions},
  author = {Poyatos, J. F. and Cirac, J. I. and Zoller, P.},
  journal = {Phys. Rev. Lett.},
  volume = {77},
  issue = {23},
  pages = {4728--4731},
  numpages = {0},
  year = {1996},
  month = {Dec},
  publisher = {American Physical Society},
  doi = {10.1103/PhysRevLett.77.4728},
  url = {https://link.aps.org/doi/10.1103/PhysRevLett.77.4728}
}

@article{Didier_18,
  title = {Remote entanglement stabilization and concentration by quantum reservoir engineering},
  author = {Didier, Nicolas and Guillaud, J\'er\'emie and Shankar, Shyam and Mirrahimi, Mazyar},
  journal = {Phys. Rev. A},
  volume = {98},
  issue = {1},
  pages = {012329},
  numpages = {8},
  year = {2018},
  month = {Jul},
  publisher = {American Physical Society},
  doi = {10.1103/PhysRevA.98.012329},
  url = {https://link.aps.org/doi/10.1103/PhysRevA.98.012329}
}

@article{kraus_04,
  title = {Discrete Entanglement Distribution with Squeezed Light},
  author = {Kraus, B. and Cirac, J. I.},
  journal = {Phys. Rev. Lett.},
  volume = {92},
  issue = {1},
  pages = {013602},
  numpages = {4},
  year = {2004},
  month = {Jan},
  publisher = {American Physical Society},
  doi = {10.1103/PhysRevLett.92.013602},
  url = {https://link.aps.org/doi/10.1103/PhysRevLett.92.013602}
}

@article{Sels_17,
author = {Dries Sels  and Anatoli Polkovnikov },
title = {Minimizing irreversible losses in quantum systems by local counterdiabatic driving},
journal = {Proc. Natl. Acad. Sci.},
volume = {114},
number = {20},
pages = {E3909-E3916},
year = {2017},
doi = {10.1073/pnas.1619826114},
}

@article{Fentaw_25,
   title={Exploring quantum control landscape and solution space complexity through optimization algorithms and dimensionality reduction},
   volume={15},
   ISSN={2045-2322},
   url={http://dx.doi.org/10.1038/s41598-025-95161-0},
   number={1},
   journal={Sci. Rep.},
   publisher={Springer Science and Business Media LLC},
   author={Fentaw, Haftu W. and Campbell, Steve and Caton, Simon},
   year={2025},
   month=apr }

@article{Moore_11,
   title={Search complexity and resource scaling for the quantum optimal control of unitary transformations},
   volume={83},
   ISSN={1094-1622},
   url={http://dx.doi.org/10.1103/PhysRevA.83.012326},
   number={1},
   journal={Phys. Rev. A},
   publisher={American Physical Society (APS)},
   author={Moore, Katharine W. and Chakrabarti, Raj and Riviello, Gregory and Rabitz, Herschel},
   year={2011},
   month=jan }

@article{Metz_2023,
   title={Self-correcting quantum many-body control using reinforcement learning with tensor networks},
   volume={5},
   ISSN={2522-5839},
   url={http://dx.doi.org/10.1038/s42256-023-00687-5},
   DOI={10.1038/s42256-023-00687-5},
   number={7},
   journal={Nat. Mach. Intell.},
   publisher={Springer Science and Business Media LLC},
   author={Metz, Friederike and Bukov, Marin},
   year={2023},
   month=jul, pages={780} }

@article{Larocca_2022,
   title={Diagnosing Barren Plateaus with Tools from Quantum Optimal Control},
   volume={6},
   ISSN={2521-327X},
   url={http://dx.doi.org/10.22331/q-2022-09-29-824},
   DOI={10.22331/q-2022-09-29-824},
   journal={Quantum},
   publisher={Verein zur Forderung des Open Access Publizierens in den Quantenwissenschaften},
   author={Larocca, Martin and Czarnik, Piotr and Sharma, Kunal and Muraleedharan, Gopikrishnan and Coles, Patrick J. and Cerezo, M.},
   year={2022},
   month=sep, pages={824} }

@article{Lu_24,
doi = {10.1088/2399-6528/ad22e5},
url = {https://doi.org/10.1088/2399-6528/ad22e5},
year = {2024},
month = {feb},
publisher = {IOP Publishing},
volume = {8},
number = {2},
pages = {025002},
author = {Lu, Yunwei and Joshi, Sandeep and San Dinh, Vinh and Koch, Jens},
title = {Optimal control of large quantum systems: assessing memory and runtime performance of GRAPE},
journal = {J. Phys. Commun.},
}

@article{Ding_26,
  author  = {Ding, Lixiang and Fan, Jingtao and Qiu, Xingze},
  title   = {Universally robust control of open quantum systems},
  journal = {npj Quantum Inf.},
  volume  = {12},
  pages   = {22},
  year    = {2026},
  doi     = {10.1038/s41534-025-01166-y}
}

@article{Chen_25,
   title={Robust and optimal control of open quantum systems},
   volume={11},
   ISSN={2375-2548},
   url={http://dx.doi.org/10.1126/sciadv.adr0875},
   number={9},
   journal={Sci. Adv.},
   publisher={American Association for the Advancement of Science (AAAS)},
   author={Chen, Zi-Jie and Huang, Hongwei and Sun, Lida and Jie, Qing-Xuan and Zhou, Jie and Hua, Ziyue and Xu, Yifang and Wang, Weiting and Guo, Guang-Can and Zou, Chang-Ling and Sun, Luyan and Zou, Xu-Bo},
   year={2025},
    pages={eadr0875},
   month=feb }

@article{Koch_22,
   title={Quantum optimal control in quantum technologies. Strategic report on current status, visions and goals for research in Europe},
   volume={9},
   ISSN={2196-0763},
   url={http://dx.doi.org/10.1140/epjqt/s40507-022-00138-x},
   number={1},
   journal={EPJ Quantum Technol.},
   publisher={Springer Science and Business Media LLC},
   author={Koch, Christiane P. and Boscain, Ugo and Calarco, Tommaso and Dirr, Gunther and Filipp, Stefan and Glaser, Steffen J. and Kosloff, Ronnie and Montangero, Simone and Schulte-Herbrüggen, Thomas and Sugny, Dominique and Wilhelm, Frank K.},
   year={2022},
   month={jul},
    pages={19},
}

@article{Schindler_13,
   title={Quantum simulation of dynamical maps with trapped ions},
   volume={9},
   ISSN={1745-2481},
   url={http://dx.doi.org/10.1038/nphys2630},
   DOI={10.1038/nphys2630},
   number={6},
   journal={Nat. Phys.},
   publisher={Springer Science and Business Media LLC},
   author={Schindler, P. and Müller, M. and Nigg, D. and Barreiro, J. T. and Martinez, E. A. and Hennrich, M. and Monz, T. and Diehl, S. and Zoller, P. and Blatt, R.},
   year={2013},
   month=may, 
    pages={361} 
}

@article{Barreiro_11,
   title={An open-system quantum simulator with trapped ions},
   volume={470},
   ISSN={1476-4687},
   url={http://dx.doi.org/10.1038/nature09801},
   DOI={10.1038/nature09801},
   number={7335},
   journal={Nature},
   publisher={Springer Science and Business Media LLC},
   author={Barreiro, Julio T. and Müller, Markus and Schindler, Philipp and Nigg, Daniel and Monz, Thomas and Chwalla, Michael and Hennrich, Markus and Roos, Christian F. and Zoller, Peter and Blatt, Rainer},
   year={2011},
   month=feb, pages={486} }

@article{Diehl_08,
   title={Quantum states and phases in driven open quantum systems with cold atoms},
   volume={4},
   ISSN={1745-2481},
   url={http://dx.doi.org/10.1038/nphys1073},
   DOI={10.1038/nphys1073},
   number={11},
   journal={Nat. Phys.},
   publisher={Springer Science and Business Media LLC},
   author={Diehl, S. and Micheli, A. and Kantian, A. and Kraus, B. and Büchler, H. P. and Zoller, P.},
   year={2008},
   month=sep, 
    pages={878} 
}

@book{Landau_76,
  author = {Landau, L. D. and Lifshitz, E. M.},
  day = 15,
  edition = 3,
  howpublished = {Paperback},
  isbn = {0750628960},
address = {Oxford},
  month = jan,
  publisher = {Butterworth-Heinemann},
  title = {Mechanics, Third Edition: Volume 1 (Course of Theoretical Physics)},
  url = {http://www.worldcat.org/isbn/0750628960},
  year = 1976
}

@article{Machnes_18,
  title = {Tunable, Flexible, and Efficient Optimization of Control Pulses for Practical Qubits},
  author = {Machnes, Shai and Ass\'emat, Elie and Tannor, David and Wilhelm, Frank K.},
  journal = {Phys. Rev. Lett.},
  volume = {120},
  issue = {15},
  pages = {150401},
  numpages = {6},
  year = {2018},
  month = {Apr},
  publisher = {American Physical Society},
  doi = {10.1103/PhysRevLett.120.150401},
  url = {https://link.aps.org/doi/10.1103/PhysRevLett.120.150401}
}

@article{Pichler_15,
  title = {Quantum optics of chiral spin networks},
  author = {Pichler, Hannes and Ramos, Tom\'as and Daley, Andrew J. and Zoller, Peter},
  journal = {Phys. Rev. A},
  volume = {91},
  issue = {4},
  pages = {042116},
  numpages = {19},
  year = {2015},
  month = {Apr},
  publisher = {American Physical Society},
  doi = {10.1103/PhysRevA.91.042116},
  url = {https://link.aps.org/doi/10.1103/PhysRevA.91.042116}
}

@misc{Yang_26,
      title={Many-body quantum optics in a cascaded chiral network}, 
      author={Frank Yang and Parth S. Shah and Chaitali Joshi and Mohammad Mirhosseini},
      year={2026},
      eprint={2607.05760},
      archivePrefix={arXiv},
      primaryClass={quant-ph}
}

@article{Diehl_11,
  title = {Topology by dissipation in atomic quantum wires},
  author = {Diehl, Sebastian and Rico, Enrique and   Baranov, Mikhail A. and Zoller, Peter},
  journal = {Nat. Phys.},
  volume = {7},
  pages = {971},
  year = {2011},
  doi = {10.1038/nphys2106},
  url = {https://doi.org/10.1038/nphys2106}
}

@article{Morigi_15,
  title = {Dissipative Quantum Control of a Spin Chain},
  author = {Morigi, Giovanna and Eschner, J\"urgen and Cormick, Cecilia and Lin, Yiheng and Leibfried, Dietrich and Wineland, David J.},
  journal = {Phys. Rev. Lett.},
  volume = {115},
  issue = {20},
  pages = {200502},
  numpages = {6},
  year = {2015},
  month = {Nov},
  publisher = {American Physical Society},
  doi = {10.1103/PhysRevLett.115.200502},
  url = {https://link.aps.org/doi/10.1103/PhysRevLett.115.200502}
}

@article{Harrington_22,
  author    = {Harrington, Patrick M. and Mueller, Erich J. and Murch, Kater W.},
  title     = {Engineered dissipation for quantum information science},
  journal   = {Nat. Rev. Phys.},
  volume    = {4},
  number    = {10},
  pages     = {660--671},
  year      = {2022},
  publisher = {Springer Science and Business Media LLC},
  doi       = {10.1038/s42254-022-00494-8},
  url = {https://doi.org/10.1038/s42254-022-00494-8}
}

@article{Kapit_17,
doi = {10.1088/2058-9565/aa7e5d},
url = {https://doi.org/10.1088/2058-9565/aa7e5d},
year = {2017},
month = {aug},
publisher = {IOP Publishing},
volume = {2},
number = {3},
pages = {033002},
author = {Kapit, Eliot},
title = {The upside of noise: engineered dissipation as a resource in superconducting circuits},
journal = {Quantum Sci. Technol.},
}

@article{Cian_19,
  title = {Photon Pair Condensation by Engineered Dissipation},
  author = {Cian, Ze-Pei and Zhu, Guanyu and Chu, Su-Kuan and Seif, Alireza and DeGottardi, Wade and Jiang, Liang and Hafezi, Mohammad},
  journal = {Phys. Rev. Lett.},
  volume = {123},
  issue = {6},
  pages = {063602},
  numpages = {6},
  year = {2019},
  month = {Aug},
  publisher = {American Physical Society},
  doi = {10.1103/PhysRevLett.123.063602},
  url = {https://link.aps.org/doi/10.1103/PhysRevLett.123.063602}
}

@article{Lin_13,
  author    = {Lin, Y. and Gaebler, J. P. and Reiter, F. and Tan, T. R. and Bowler, R. and S{\o}rensen, A. S. and Leibfried, D. and Wineland, D. J.},
  title     = {Dissipative production of a maximally entangled steady state of two quantum bits},
  journal   = {Nature},
  volume    = {504},
  number    = {7480},
  pages     = {415--418},
  year      = {2013},
  publisher = {Nature Publishing Group},
  doi       = {10.1038/nature12801}
}

@misc{Wang_10,
  author        = {Wang, Xiaoting and Schirmer, Sophie G.},
  title         = {Generating maximal entanglement between non-interacting atoms by collective decay and symmetry breaking},
  year          = {2010},
  eprint        = {1005.2114},
  archiveprefix = {arXiv},
  primaryclass  = {quant-ph}
}

@article{Reiter_16,
  title = {Scalable Dissipative Preparation of Many-Body Entanglement},
  author = {Reiter, Florentin and Reeb, David and S\o{}rensen, Anders S.},
  journal = {Phys. Rev. Lett.},
  volume = {117},
  issue = {4},
  pages = {040501},
  numpages = {6},
  year = {2016},
  month = {Jul},
  publisher = {American Physical Society},
  doi = {10.1103/PhysRevLett.117.040501},
  url = {https://link.aps.org/doi/10.1103/PhysRevLett.117.040501}
}

@article{Rao_14,
  title = {Deterministic entanglement of Rydberg ensembles by engineered dissipation},
  author = {Rao, D. D. Bhaktavatsala and M\o{}lmer, Klaus},
  journal = {Phys. Rev. A},
  volume = {90},
  issue = {6},
  pages = {062319},
  numpages = {5},
  year = {2014},
  month = {Dec},
  publisher = {American Physical Society},
  doi = {10.1103/PhysRevA.90.062319},
  url = {https://link.aps.org/doi/10.1103/PhysRevA.90.062319}
}

@article{Aron_14,
  title = {Steady-state entanglement of spatially separated qubits via quantum bath engineering},
  author = {Aron, Camille and Kulkarni, Manas and T\"ureci, Hakan E.},
  journal = {Phys. Rev. A},
  volume = {90},
  issue = {6},
  pages = {062305},
  numpages = {5},
  year = {2014},
  month = {Dec},
  publisher = {American Physical Society},
  doi = {10.1103/PhysRevA.90.062305},
  url = {https://link.aps.org/doi/10.1103/PhysRevA.90.062305}
}

@article{Shankar_13,
  author    = {Shankar, S. and Hatridge, M. and Leghtas, Z. and Sliwa, K. M. and Narla, A. and Vool, U. and Girvin, S. M. and Frunzio, L. and Devoret, M. H. and Schoelkopf, R. J.},
  title     = {Autonomously stabilized entanglement between two superconducting quantum bits},
  journal   = {Nature},
  volume    = {504},
  number    = {7480},
  pages     = {419--422},
  year      = {2013},
  publisher = {Nature Publishing Group},
  doi       = {10.1038/nature12802},
 url = {https://doi.org/10.1038/nphys2106}
}

@article{Cole_22,
  title = {Resource-Efficient Dissipative Entanglement of Two Trapped-Ion Qubits},
  author = {Cole, Daniel C. and Erickson, Stephen D. and Zarantonello, Giorgio and Horn, Karl P. and Hou, Pan-Yu and Wu, Jenny J. and Slichter, Daniel H. and Reiter, Florentin and Koch, Christiane P. and Leibfried, Dietrich},
  journal = {Phys. Rev. Lett.},
  volume = {128},
  issue = {8},
  pages = {080502},
  numpages = {7},
  year = {2022},
  month = {Feb},
  publisher = {American Physical Society},
  doi = {10.1103/PhysRevLett.128.080502},
  url = {https://link.aps.org/doi/10.1103/PhysRevLett.128.080502}
}

@article{Kastoryano_11,
  title = {Dissipative Preparation of Entanglement in Optical Cavities},
  author = {Kastoryano, M. J. and Reiter, F. and S\o{}rensen, A. S.},
  journal = {Phys. Rev. Lett.},
  volume = {106},
  issue = {9},
  pages = {090502},
  numpages = {4},
  year = {2011},
  month = {Feb},
  publisher = {American Physical Society},
  doi = {10.1103/PhysRevLett.106.090502},
  url = {https://link.aps.org/doi/10.1103/PhysRevLett.106.090502}
}

@article{Benatti_03,
  title = {Environment Induced Entanglement in Markovian Dissipative Dynamics},
  author = {Benatti, Fabio and Floreanini, Roberto and Piani, Marco},
  journal = {Phys. Rev. Lett.},
  volume = {91},
  issue = {7},
  pages = {070402},
  numpages = {4},
  year = {2003},
  month = {Aug},
  publisher = {American Physical Society},
  doi = {10.1103/PhysRevLett.91.070402},
  url = {https://link.aps.org/doi/10.1103/PhysRevLett.91.070402}
}

@article{Paternostro_04,
  title = {Complete Conditions for Entanglement Transfer},
  author = {Paternostro, M. and Son, W. and Kim, M. S.},
  journal = {Phys. Rev. Lett.},
  volume = {92},
  issue = {19},
  pages = {197901},
  numpages = {4},
  year = {2004},
  month = {May},
  publisher = {American Physical Society},
  doi = {10.1103/PhysRevLett.92.197901},
  url = {https://link.aps.org/doi/10.1103/PhysRevLett.92.197901}
}

@article{Felicetti_14,
  title = {Dynamical Casimir Effect Entangles Artificial Atoms},
  author = {Felicetti, S. and Sanz, M. and Lamata, L. and Romero, G. and Johansson, G. and Delsing, P. and Solano, E.},
  journal = {Phys. Rev. Lett.},
  volume = {113},
  issue = {9},
  pages = {093602},
  numpages = {5},
  year = {2014},
  month = {Aug},
  publisher = {American Physical Society},
  doi = {10.1103/PhysRevLett.113.093602},
  url = {https://link.aps.org/doi/10.1103/PhysRevLett.113.093602}
}

@article{You_18,
  title = {Waveguide quantum electrodynamics in squeezed vacuum},
  author = {You, Jieyu and Liao, Zeyang and Li, Sheng-Wen and Zubairy, M. Suhail},
  journal = {Phys. Rev. A},
  volume = {97},
  issue = {2},
  pages = {023810},
  numpages = {15},
  year = {2018},
  month = {Feb},
  publisher = {American Physical Society},
  doi = {10.1103/PhysRevA.97.023810},
  url = {https://link.aps.org/doi/10.1103/PhysRevA.97.023810}
}

@article{Ma_19,
  title = {Stabilizing Bell states of two separated superconducting qubits via quantum reservoir engineering},
  author = {Ma, Sheng-li and Li, Xin-ke and Liu, Xin-yu and Xie, Ji-kun and Li, Fu-li},
  journal = {Phys. Rev. A},
  volume = {99},
  issue = {4},
  pages = {042336},
  numpages = {6},
  year = {2019},
  month = {Apr},
  publisher = {American Physical Society},
  doi = {10.1103/PhysRevA.99.042336},
  url = {https://link.aps.org/doi/10.1103/PhysRevA.99.042336}
}

@article{Zippilli_13,
  title = {Entanglement Replication in Driven Dissipative Many-Body systems},
  author = {Zippilli, S. and Paternostro, M. and Adesso, G. and Illuminati, F.},
  journal = {Phys. Rev. Lett.},
  volume = {110},
  issue = {4},
  pages = {040503},
  numpages = {5},
  year = {2013},
  month = {Jan},
  publisher = {American Physical Society},
  doi = {10.1103/PhysRevLett.110.040503},
  url = {https://link.aps.org/doi/10.1103/PhysRevLett.110.040503}
}

@article{Pocklington_22,
  title = {Stabilizing volume-law entangled states of fermions and qubits using local dissipation},
  author = {Pocklington, Andrew and Wang, Yu-Xin and Yanay, Yariv and Clerk, A. A.},
  journal = {Phys. Rev. B},
  volume = {105},
  issue = {14},
  pages = {L140301},
  numpages = {6},
  year = {2022},
  month = {Apr},
  publisher = {American Physical Society},
  doi = {10.1103/PhysRevB.105.L140301},
  url = {https://link.aps.org/doi/10.1103/PhysRevB.105.L140301}
}

@article{Ramos_14,
  title = {Quantum Spin Dimers from Chiral Dissipation in Cold-Atom Chains},
  author = {Ramos, Tom\'as and Pichler, Hannes and Daley, Andrew J. and Zoller, Peter},
  journal = {Phys. Rev. Lett.},
  volume = {113},
  issue = {23},
  pages = {237203},
  numpages = {6},
  year = {2014},
  month = {Dec},
  publisher = {American Physical Society},
  doi = {10.1103/PhysRevLett.113.237203},
  url = {https://link.aps.org/doi/10.1103/PhysRevLett.113.237203}
}

@article{Muschik_11,
  title = {Dissipatively driven entanglement of two macroscopic atomic ensembles},
  author = {Muschik, Christine A. and Polzik, Eugene S. and Cirac, J. Ignacio},
  journal = {Phys. Rev. A},
  volume = {83},
  issue = {5},
  pages = {052312},
  numpages = {19},
  year = {2011},
  month = {May},
  publisher = {American Physical Society},
  doi = {10.1103/PhysRevA.83.052312},
  url = {https://link.aps.org/doi/10.1103/PhysRevA.83.052312}
}

@misc{Almanakly_26,
  author        = {Almanakly, Aziza and Soro, Ariadna and Vivas-Via{\~n}a, Alejandro and Yankelevich, Beatriz and Groiseau, Caspar and Pahl, David and An, Junyoung and Cutter, Gabriel and Gingras, Michael E. and Niedzielski, Bethany M. and Stickler, Hannah and DeP{\'e}ncier Pi{\~n}ero, Ren{\'e}e and Schwartz, Mollie E. and Serniak, Kyle and Hays, Max and Grover, Jeffrey A. and Kockum, Anton Frisk and Oliver, William D.},
  title         = {Driven-dissipative entanglement of distant giant atoms},
  journal       = {arXiv preprint arXiv:2606.13375},
  year          = {2026},
  eprint        = {2606.13375},
  archiveprefix = {arXiv},
  primaryclass  = {quant-ph}
}

@article{Brown_22,
  author  = {Brown, Tristan and Doucet, Emery and Rist{\`e}, Diego and Ribeill, Guilhem and Cicak, Katarina and Aumentado, Joe and Simmonds, Ray and Govia, Luke and Kamal, Archana and Ranzani, Leonardo},
  title   = {Trade off-free entanglement stabilization in a superconducting qutrit-qubit system},
  journal = {Nat. Commun.},
  year    = {2022},
  volume  = {13},
  pages   = {3994},
  doi     = {10.1038/s41467-022-31638-0},
  url     = {https://doi.org/10.1038/s41467-022-31638-0}
}

@article{Riste_13,
  author  = {Rist{\`e}, D. and Dukalski, M. and Watson, C. A. and de Lange, G. and Tiggelman, M. J. and Blanter, Ya. M. and Lehnert, K. W. and Schouten, R. N. and DiCarlo, L.},
  title   = {Deterministic entanglement of superconducting qubits by parity measurement and feedback},
  journal = {Nature},
  year    = {2013},
  volume  = {502},
  number  = {7471},
  pages   = {350--354},
  doi     = {10.1038/nature12513},
  url     = {https://doi.org/10.1038/nature12513}
}

@article{Roch_14,
  title = {Observation of Measurement-Induced Entanglement and Quantum Trajectories of Remote Superconducting Qubits},
  author = {Roch, N. and Schwartz, M. E. and Motzoi, F. and Macklin, C. and Vijay, R. and Eddins, A. W. and Korotkov, A. N. and Whaley, K. B. and Sarovar, M. and Siddiqi, I.},
  journal = {Phys. Rev. Lett.},
  volume = {112},
  issue = {17},
  pages = {170501},
  numpages = {5},
  year = {2014},
  month = {Apr},
  publisher = {American Physical Society},
  doi = {10.1103/PhysRevLett.112.170501},
  url = {https://link.aps.org/doi/10.1103/PhysRevLett.112.170501}
}

@article{Andersen_19,
  author  = {Andersen, Christian Kraglund and Remm, Ants and Balasiu, Stefania and Krinner, Sebastian and Heinsoo, Johannes and Besse, Jean-Claude and Gabureac, Mihai and Wallraff, Andreas and Eichler, Christopher},
  title   = {Entanglement stabilization using ancilla-based parity detection and real-time feedback in superconducting circuits},
  journal = {npj Quantum Inf.},
  year    = {2019},
  volume  = {5},
  pages   = {69},
  doi     = {10.1038/s41534-019-0185-4},
  url     = {https://doi.org/10.1038/s41534-019-0185-4}
}

@ARTICLE{Demirplak_2003,
       author = {{Demirplak}, Mustafa and {Rice}, Stuart A.},
        title = "{Adiabatic Population Transfer with Control Fields}",
      journal = {J. Phys. Chem. A},
         year = 2003,
        month = nov,
       volume = {107},
       number = {46},
        pages = {9937-9945},
          doi = {10.1021/jp030708a},
       adsurl = {https://ui.adsabs.harvard.edu/abs/2003JPCA..107.9937D}
}

@article{Rezakhani_09,
  title = {Quantum Adiabatic Brachistochrone},
  author = {Rezakhani, A. T. and Kuo, W.-J. and Hamma, A. and Lidar, D. A. and Zanardi, P.},
  journal = {Phys. Rev. Lett.},
  volume = {103},
  issue = {8},
  pages = {080502},
  numpages = {4},
  year = {2009},
  month = {Aug},
  publisher = {American Physical Society},
  doi = {10.1103/PhysRevLett.103.080502},
  url = {https://link.aps.org/doi/10.1103/PhysRevLett.103.080502}
}

@article{Campo_12,
  title = {Assisted Finite-Rate Adiabatic Passage Across a Quantum Critical Point: Exact Solution for the Quantum Ising Model},
  author = {del Campo, Adolfo and Rams, Marek M. and Zurek, Wojciech H.},
  journal = {Phys. Rev. Lett.},
  volume = {109},
  issue = {11},
  pages = {115703},
  numpages = {5},
  year = {2012},
  month = {Sep},
  publisher = {American Physical Society},
  doi = {10.1103/PhysRevLett.109.115703},
  url = {https://link.aps.org/doi/10.1103/PhysRevLett.109.115703}
}

@article{Li2012,
  title = {Dissipative preparation of entangled states between two spatially separated nitrogen-vacancy centers},
  author = {Li, Peng-Bo and Gao, Shao-Yan and Li, Hong-Rong and Ma, Sheng-Li and Li, Fu-Li},
  journal = {Phys. Rev. A},
  volume = {85},
  issue = {4},
  pages = {042306},
  numpages = {7},
  year = {2012},
  month = {Apr},
  publisher = {American Physical Society},
  doi = {10.1103/PhysRevA.85.042306},
  url = {https://link.aps.org/doi/10.1103/PhysRevA.85.042306}
}

@article{Horn2018,
author = {Karl P. Horn and Florentin Reiter and Yiheng Lin and Dietrich Leibfried and Christiane P. Koch},
title = {Quantum optimal control of the dissipative production of a maximally entangled state},
journal = {New J. Phys.},
volume = {20},
number = {12},
pages = {123010},
year = {2018},
doi = {10.1088/1367-2630/aaf360},
}

@article{Eichler2011,
  title = {Observation of Two-Mode Squeezing in the Microwave Frequency Domain},
  author = {Eichler, C. and Bozyigit, D. and Lang, C. and Baur, M. and Steffen, L. and Fink, J. M. and Filipp, S. and Wallraff, A.},
  journal = {Phys. Rev. Lett.},
  volume = {107},
  issue = {11},
  pages = {113601},
  numpages = {5},
  year = {2011},
  month = {Sep},
  publisher = {American Physical Society},
  doi = {10.1103/PhysRevLett.107.113601},
  url = {https://link.aps.org/doi/10.1103/PhysRevLett.107.113601}
}

@article{Menzel2012,
  title = {Path Entanglement of Continuous-Variable Quantum Microwaves},
  author = {Menzel, E. P. and Di Candia, R. and Deppe, F. and Eder, P. and Zhong, L. and Ihmig, M. and Haeberlein, M. and Baust, A. and Hoffmann, E. and Ballester, D. and Inomata, K. and Yamamoto, T. and Nakamura, Y. and Solano, E. and Marx, A. and Gross, R.},
  journal = {Phys. Rev. Lett.},
  volume = {109},
  issue = {25},
  pages = {250502},
  numpages = {4},
  year = {2012},
  month = {Dec},
  publisher = {American Physical Society},
  doi = {10.1103/PhysRevLett.109.250502},
  url = {https://link.aps.org/doi/10.1103/PhysRevLett.109.250502}
}

@article{Flurin2012,
  title = {Generating Entangled Microwave Radiation Over Two Transmission Lines},
  author = {Flurin, E. and Roch, N. and Mallet, F. and Devoret, M. H. and Huard, B.},
  journal = {Phys. Rev. Lett.},
  volume = {109},
  issue = {18},
  pages = {183901},
  numpages = {5},
  year = {2012},
  month = {Oct},
  publisher = {American Physical Society},
  doi = {10.1103/PhysRevLett.109.183901},
  url = {https://link.aps.org/doi/10.1103/PhysRevLett.109.183901}
}

@article{Couteau2018,
author = {Christophe Couteau},
title = {Spontaneous parametric down-conversion},
journal = {Cont. Phys.},
volume = {59},
number = {3},
pages = {291--304},
year = {2018},
publisher = {Taylor \& Francis},
doi = {10.1080/00107514.2018.1488463},
URL = {https://doi.org/10.1080/00107514.2018.1488463}
}
    
    \end{document}